\RequirePackage{fix-cm}
\documentclass[smallextended]{article}       
\usepackage{natbib}
\usepackage{subcaption}
\usepackage{amsmath}
\usepackage{graphicx}
\usepackage{adjustbox}
\usepackage{url} 
\usepackage{enumerate}
\usepackage{booktabs,amsfonts}
\usepackage{graphicx,psfrag,epsf}
\usepackage{epstopdf}
\usepackage{amsmath}
\usepackage{amssymb}
\usepackage[maxfloats=36]{morefloats}
\usepackage{float}
\newtheorem{theorem}{Proposition}
\usepackage{booktabs}
\usepackage[english]{babel}
\usepackage[titletoc,title]{appendix}
\usepackage{xcolor}
\usepackage{upquote}
\usepackage{bmpsize}
\usepackage{color,soul} 
\usepackage{color}
\usepackage{amssymb}
\usepackage[colorlinks]{hyperref}
\hypersetup{
  colorlinks=true,
  citecolor=blue,
  linkcolor=red,
  }
\usepackage{authblk}

\usepackage{graphicx}
\begin{document}

\title{Multivariate outlier detection based on a robust Mahalanobis distance with shrinkage estimators}

\author[1]{Elisa Cabana\thanks{This research was partially supported by Spanish Ministry grant ECO2015-66593-P.}}
\author[2]{Rosa E. Lillo}
\author[3]{Henry Laniado}

\affil[1]{Department of Statistics, Universidad Carlos III de Madrid, Spain.}
\affil[2]{Department of Statistics and UC3M-Santander Big Data Institute, Universidad Carlos III de Madrid, Spain.}
\affil[3]{Department of Mathematical Sciences,  Universidad EAFIT, Medell\'in, Colombia.}
\date{}                     
\setcounter{Maxaffil}{0}
\renewcommand\Affilfont{\itshape\small}

\maketitle

\begin{abstract}
A collection of robust Mahalanobis distances for multivariate outlier detection is proposed, based on the notion of shrinkage. Robust intensity and scaling factors are optimally estimated to define the shrinkage. Some properties are investigated, such as affine equivariance and breakdown value. The performance of the proposal is illustrated through the comparison to other techniques from the literature, in a simulation study and with a real dataset. The behavior when the underlying distribution is heavy-tailed or skewed, shows the appropriateness of the method when we deviate from the common assumption of normality. The resulting high correct detection rates and low false detection rates in the vast majority of cases, as well as the significantly smaller computation time shows the advantages of our proposal.
\end{abstract}

{\it Keywords:}  multivariate distance,  robust location and covariance matrix estimation, comedian matrix, multivariate $L_1$-median.
\vfill
\hfill {}

\section{Introduction}
\label{intro}
The detection of outliers in multivariate data is an important task in Statistics, since that kind of data can distort any statistical procedure (\cite{tarr2016robust}). The task of detecting multivariate outliers can be useful in various fields such as quality control, medicine, finance, image analysis, and chemistry (\cite{vargas2003robust}, \cite{brettschneider2008quality}, \cite{hubert2008high}, \cite{hubert2010minimum},  \cite{perrotta2010detecting} and \cite{choi2016multivariate}). The concept of outlier is not well-defined in the literature since different authors tend to have varying definitions. Although they are generally defined as observations resulting from a secondary process, which differ from the background distribution. Thus, the outliers will differ from the main bulk of the data. They do not need to be especially high or low for all the variables in the data set, that is why the task of identifying multivariate outliers with the classical univariate methods commonly fail. In the multivariate sense, there must be considered both the distance of an observation from the centroid of the data, and the shape of the data. The Mahalanobis distance \citep{mahalanobis1936generalized} is a well-known  measure which takes it into account. For multivariate gaussian data, the distribution of the squared Mahalanobis distance, $MD^2$, is known \citep{gnanadesikan1972robust} to be chi-squared with $p$ (the dimension of the data, the number of variables) degrees of freedom, i.e. $\chi_p^2$. Then, the adopted rule for identifying the outliers is selecting the threshold as the 0.975 quantile of the $\chi_p^2$.

However, outliers need not necessarily have large MD values (Masking problem) and not all observations with large MD values are necessarily outliers (Swamping problem) \citep{hadi1992identifying}. The problems of masking and swamping arise due to the influence of outliers on classical location and scatter estimates (sample mean and sample covariance matrix), which implies that the estimated distance will not be robust to outliers. The solution is to consider robust estimators of centrality and covariance matrix to obtain a robust Mahalanobis distance (\textit{RMD}). Many robust estimators for location and covariance have been introduced in the literature \citep{maronna1976robust}. \cite{rousseeuw1985multivariate} proposed the Minimum Covariance Determinant (\textit{MCD}) estimator based on the computation of the ellipsoid with the smallest volume or with the smallest covariance determinant that would encompass at least half of the data points. The procedure required naive subsampling for minimizing the objective function of the \textit{MCD}, but an improvement much more effective, the \textit{Fast-MCD}, was introduced by \cite{rousseeuw1999fast} and a code is available in MATLAB (\cite{verboven2005libra}). Unfortunately \textit{Fast-MCD}
still requires substantial running times for large $p$, because the number of candidate solutions grows exponentially with the dimension $p$ of the sample and, as a consequence, the procedure becomes computationally expensive for even moderately sized problems. 
 
On the other hand, the squared \textit{RMD} distributional fit usually breaks down, i.e. it does not necessarily have to follow a chi-squared distribution when you deviate from the gaussian distribution. Thus, determining exact cutoff values for outlying distances continues to be a difficult problem and it has found much attention because no universally applicable method has been proposed. Despite this fact, the $\chi^2_{p;0.975}$ quantile is often considered as threshold for recognizing outliers in the robust distance case, but this approach may have some drawbacks. Evidence of this behavior is now well documented even in moderately large samples, especially when the number of variables increases (\cite{becker1999masking}, \cite{hardin2005distribution}, \cite{cerioli2009controlling} and \cite{atkinson2008fitting}). It is crucial to determine the threshold for the distances in order to decide whether an observation is an outlier. \cite{filzmoser2005multivariate} proposed to use an adjusted quantile, instead of the classical choice of the $\chi^2_{p;0.975}$ quantile. The adjusted threshold is estimated adaptively from the data, but their proposal is defined for a specific robust Mahalanobis distance, the one based on the \textit{MCD} estimator. Let us call this method \textit{Adj MCD}. \cite{pena2001multivariate} and \cite{pena2007combining} proposed an algorithm called \textit{Kurtosis}, based on the analysis of the projections of the sample points onto a certain set of directions obtained by maximizing and minimizing the kurtosis coefficient of the projections, and some random directions generated by a stratified sampling scheme. With the combination of random
and specific directions, the authors proposed a powerful procedure for robust estimation and outlier detection. However, this procedure has some drawbacks when the dimension $p$ of the sample space grows, and in presence of correlation between the variables, the method looses power \citep{marcano2013comparacion}. \cite{maronna2002robust} proposed the Orthogonalized Gnanadesikan-Kettenring (\textit{OGK}) estimator. It was the result of applying a general method to the pairwise robust scatter matrix from \cite{gnanadesikan1972robust}, in order to obtain a positive-definite scatter matrix. On the other hand, a reweighing step can be used to identify  outliers, where atypical observations get weight 0 and normal observations get weight 1. \cite{sajesh2012outlier} proposed the Comedian method (\textit{COM}) to detect outliers from multivariate
data based on the comedian matrix estimator from \cite{falk1997mad}. The method is found to be efficient under various simulation scenarios and suitable in high-dimensional data. Furthermore, there are several real scenarios where the number of variables is high in which outlier detection is very important. For example, medical imaging datasets often contain deviant observations due to pre-processing artifacts or large intrinsic inter-subject variability (\cite{lazar2008statistical}, \cite{lindquist2008statistical}, \cite{monti2011statistical}, \cite{poline2012general}), in biological and financial studies  (\cite{chen2010two} and \cite{zeng2015aberrant}), and also in geochemical data, because of their complex nature (\cite{reimann2000normal}, \cite{templ2008cluster}).

In this article, a collection of \textit{RMD's} are proposed for outlier detection especially in high dimension. They are based on considering different combinations of robust estimators of location and covariance matrix. Two basic options are considered for the location parameter: a component-wise median and the $L_1$  multivariate median (\cite{gower1974algorithm}, \cite{brown1983statistical}, \cite{dodge1987introduction}, \cite{small1990survey}). A notion called \textit{shrinkage estimator} (\cite{ledoit2003honey}, \cite{ledoit2003improved}, \cite{ledoit2004well}, \cite{demiguel2013size}, \cite{gao2016flexible}, cite{sun2018portfolio}, \cite{steland2018shrinkage}) is considered, which is aimed to reduce estimation error. The shrinkage is applied to both of the previous mentioned location estimators. As for the covariance matrix, the options basically consists on a shrinkage estimator over special cases of \textit{comedian matrices} (\cite{hall1985limit}, \cite{falk1997mad}), which are based on a location parameter that will be estimated using a robust estimator of centrality in a way that a \textit{RMD} can be obtained with meaningful combinations of both location and covariance matrix estimators. Simulation results demonstrates the satisfactory practical performance of our proposal, especially when the number of variables grows. The computational cost is studied by both simulations and a real dataset example.

The paper is organized as follows. Section 2 describes the shrinkage estimators both for the location and the covariance matrix, and the proposed combinations of these estimators in order to define a \textit{RMD}. Section 3 shows a simulation study with contaminated multivariate gaussian data and when we deviate from the gaussian assumption, e.g. with skewed or heavy-tailed data, to compare the proposal with the other robust approaches: \textit{MCD}, \textit{Adj MCD}, \textit{Kurtosis}, \textit{OGK} and \textit{COM}. To investigate the properties of affine equivariance and breakdown value, other simulation scenarios are proposed with correlated data, transformed data and large contaminated data. Section 4 shows the behavior with a real dataset example. Finally, Section 5 provides some conclusions.

\section{A robust Mahalanobis distance based on shrinkage estimator}

The classical Mahalanobis distance is defined for every $p-$dimensional observation $\mathbf{x}_{i}$ of the multivariate sample $\{ \mathbf{x}_{1},...,\mathbf{x}_{n}  \}$, as:
\begin{equation}\label{eq:maha}
MD_i=((\mathbf{x}_{i}-\hat{\boldsymbol{\mu}})\hat{\Sigma}^{-1}(\mathbf{x}_{i}-\hat{\boldsymbol{\mu}})^T)^{1/2},
\end{equation}
\noindent
where $\hat{\boldsymbol{\mu}}$ is the estimated multivariate location (sample mean) and $\hat{\Sigma}$ is the estimated covariance matrix (sample covariance matrix).

Since the problem with this definition is that the classical estimates of location and covariance matrix are often highly influenced by the presence of outliers \citep{rousseeuw1990unmasking}, the solution is to consider robust estimates of centrality and covariance matrix, i.e. resistant against the influence of outlying observations, giving rise to a robust Mahalanobis distance, defined as:
\begin{equation}\label{eq:robustmaha}
RMD_i:=((\mathbf{x}_{i}-\hat{\boldsymbol{\mu}}_R)\hat{\Sigma}_R^{-1}(\mathbf{x}_{i}-\hat{\boldsymbol{\mu}}_R)^T)^{1/2},
\end{equation}
\noindent
where $\hat{\boldsymbol{\mu}}_R$ and $\hat{\Sigma}_R$ are robust estimators of centrality and covariance matrix, respectively.

We propose to use a notion which is frequently used in finance and portfolio optimization, known as \textit{shrinkage} (Equation \ref{shrinkgeneral}). It is widely used in those fields because its good performance for ``large $p$ small $n$'' problems (see \cite{couillet2014large}, \cite{chen2011robust} and \cite{steland2018shrinkage}), although we focus on data with $n>p$. This estimator $\hat{E}_{Sh}$ relies on the fact that ``shrinking'' an estimator $\hat{E}$ of a parameter $\theta$ towards a target estimator $\hat{T}$, would help to reduce the estimation error, because although the shrinkage target is usually biased, it also contains less variance than the estimator $\hat{E}$. Therefore,  under general conditions, there exists a \textit{shrinkage intensity} $\eta$, so the resulting shrinkage estimator would contain less estimation error than $\hat{E}$ \citep{james1961estimation}.
\begin{equation}\label{shrinkgeneral}
\hat{E}_{Sh}=(1-\eta) \hat{E} + \eta \hat{T}
\end{equation}

The main advantage of using a shrinkage estimator is to obtain a trade-off between bias and variance. This approach can be applied to estimate both the location and dispersion parameters obtaining different meaningful combinations  to define robust Mahalanobis distances. In the case of covariance matrices shrinkage has the additional advantage that it is always positive definite and well conditioned.

\subsection{Location parameter}\label{subsectionlocation}

Let $\textbf{x}=\{\mathbf{x}_{\cdot1},...,\mathbf{x}_{\cdot p} \}$ be the $n \times p$ data matrix with $n$ being the sample size and $p$ the number of variables. Based on the fact that the $median$ is a better choice in terms of robustness, we start by considering as a location estimator the \textit{component-wise median:} 
\begin{equation}\label{muCCM}
\hat{\boldsymbol{\mu}}_{CCM}=(median(\mathbf{x}_{\cdot 1}),...,median(\mathbf{x}_{\cdot p})),
\end{equation}
\noindent
where $median$ denotes the univariate median and $(\mathbf{x}_{\cdot j} )= (x_{1 j},  ..., x_{n j})^T$ for all $j=1,...,p$ is the $j$-th column of $\mathbf{x}$.

Another option is to consider a multivariate median $\hat{\boldsymbol{\mu}}_{MM}$ called \textit{$L_1-$median} which is a robust and highly efficient estimator of central tendency (\cite{lopuhaa1991breakdown}, \cite{vardi2000multivariate}, \cite{oja2010multivariate}). It is defined as:
\begin{equation}\label{muMM}
\hat{\boldsymbol{\mu}}_{MM}=argmin_{\mathbf{x}_{m},  \ m \in \{1,...,n\}} \frac{1}{n} \sum_{i=1}^n ||\mathbf{x}_{m} - \mathbf{x}_{i}||_1
\end{equation}

\cite{demiguel2013size} proposed a shrinkage estimator over the sample mean, towards a scaled vector of ones as the target. In the same way we propose to study shrinkage estimators for both (\ref{muCCM}) and (\ref{muMM}). Consider $\nu_{\boldsymbol{\mu}} \mathbf{e}$ as the target estimator $\hat{T}$ in (\ref{shrinkgeneral}), where $\mathbf{e}$ is the $p-$dimensional vector of ones, and consider $\hat{\boldsymbol{\mu}}_{CCM}$ as the sample estimator $\hat{E}$. Then, the shrinkage estimator over the component-wise median is: 
\begin{equation}\label{ShrinkagetoMuCCM}
\hat{\boldsymbol{\mu}}_{Sh(CCM)}=(1- \eta) \hat{\boldsymbol{\mu}}_{CCM} + \eta \nu_{\boldsymbol{\mu}} \mathbf{e}
\end{equation}

The scaling factor $\nu_{\boldsymbol{\mu}}$ and the intensity $\eta$ should minimize the expected quadratic loss, that is:
\begin{equation}\label{eq:minsquarederrormedianCCM}
\begin{matrix}
\textup{min}_{\nu_{\boldsymbol{\mu}},\eta} & E\left [ \left \| \boldsymbol{\hat{\mu}}_{Sh(CCM)} - \boldsymbol{\mu}  \right \|_2^2 \right ]\\ 
\textup{s.t.} & \boldsymbol{\hat{\mu}}_{Sh(CCM)} = (1-\eta) \boldsymbol{\hat{\mu}}_{CCM} + \eta \nu_{\boldsymbol{\mu}} \mathbf{e},
\end{matrix}
\end{equation}
\noindent
where $\left \| \mathbf{x} \right \|_2^2=\sum_{j=1}^p x_j^2$.

\begin{theorem}\label{prop1teo}
The solution of the problem in (\ref{eq:minsquarederrormedianCCM}) is:
\begin{align}
\nu_{\boldsymbol{\mu}}&=\frac{\hat{\boldsymbol{\mu}}_{CCM} \mathbf{e} }{p}, \quad \eta= \frac{E\left [ \left \| \hat{\boldsymbol{\mu}}_{CCM}-\boldsymbol{\mu} \right \|_2^2 \right ]}{E\left [ \left \| \hat{\boldsymbol{\mu}}_{CCM}- \nu_{\boldsymbol{\mu}}\mathbf{e} \right \|_2^2 \right ]  }\label{alphaprop1}
\end{align}
\end{theorem}

See the proof in Section 1 from the Supplementary Material. Note that the denominator in the above expression (\ref{alphaprop1}) has no problem when estimating, but the numerator is not straightforward because $\boldsymbol{\mu}$ is unknown. Then, it is necessary to provide another expression for the numerator. \cite{chu1955distribution} investigated the distribution for the sample median estimator and obtained the following result about the variance in presence of normality. Fix $j$, for $j \in \{1,...,p\}$:
\begin{equation}
\sigma^2_{{{\boldsymbol{\hat{\mu}}}_{CCM}}_j}=Var({{\boldsymbol{\hat{\mu}}}_{CCM}}_j)=\frac{\pi}{2n} \sigma_{\mathbf{x_{\cdot j}}}^2
\end{equation}
Therefore, the numerator in the expression (\ref{alphaprop1}) for determining the $\eta$ in Proposition \ref{prop1teo} is:
\begin{align}\label{ExpectationmuCCM}
 E\left [ \left \| {\hat{\boldsymbol{\mu}}}_{CCM}-\boldsymbol{\mu} \right \|_2^2 \right ] &= E\left [ \sum_{j=1}^p ({{\boldsymbol{\hat{\mu}}}_{CCM}}_j-\boldsymbol{\mu}_j )^2 \right]=\sum_{j=1}^p   \sigma^2_{{{\boldsymbol{\hat{\mu}}_{CCM}}_j}} =   \frac{\pi}{2n} \sum_{j=1}^p \sigma_{\mathbf{x}_{\cdot j}}^2 
\end{align}

We need to estimate $\sigma_{\mathbf{x}_{\cdot j}}^2$ robustly and we will do so as explained in the next sub-section with property (\ref{sigmaCCM}).

On the other hand, consider $\nu_{\boldsymbol{\mu}} \mathbf{e}$ again as the target estimator $\hat{T}$ and consider  $\hat{\boldsymbol{\mu}}_{MM}$ as the sample estimator $\hat{E}$, in (\ref{shrinkgeneral}). Then, the shrinkage estimator over the multivariate $L_1-$median is:
\begin{equation}\label{shrinkmuMM}
\hat{\boldsymbol{\mu}}_{Sh(MM)}=(1- \eta) \hat{\boldsymbol{\mu}}_{MM} + \eta \nu_{\boldsymbol{\mu}} \mathbf{e} 
\end{equation}

The scaling factor $\nu_{\boldsymbol{\mu}}$ and the intensity $\eta$ should minimize the expected quadratic loss:
\begin{equation}\label{eq:minsquarederrormedianMM}
\begin{matrix}
\textup{min}_{\nu_{\boldsymbol{\mu}},\eta} & E\left [ \left \| \boldsymbol{\hat{\mu}}_{Sh(MM)} - \boldsymbol{\mu}  \right \|_2^2 \right ]\\ 
\textup{s.t.} & \boldsymbol{\hat{\mu}}_{Sh(MM)} = (1-\eta) \boldsymbol{\hat{\mu}}_{MM} + \eta \nu_{\boldsymbol{\mu}} \mathbf{e},
\end{matrix}
\end{equation}
\noindent
where $\left \| \mathbf{x} \right \|_2^2=\sum_{j=1}^p x_j^2$.

\begin{theorem}\label{prop2teo}

The solution of the problem in (\ref{eq:minsquarederrormedianMM}) is: 
\begin{align}
\nu_{\boldsymbol{\mu}}&=\frac{\hat{\boldsymbol{\mu}}_{MM}\mathbf{e} }{p}, \quad
\eta= \frac{E\left [ \left \| \hat{\boldsymbol{\mu}}_{MM}-\boldsymbol{\mu} \right \|^2 \right ]}{E\left [ \left \| \hat{\boldsymbol{\mu}}_{MM}- \nu_{\boldsymbol{\mu}}\mathbf{e}  \right \|^2 \right ]  }\label{alphaprop2}
\end{align}
\end{theorem}

The proof is also in Section 1 from the Supplementary Material. As in the previous case, the denominator in the $\eta$ expression (\ref{alphaprop2}) can be described as:
\begin{equation}\label{eq:forboot}
 E\left [ \left \| {\hat{\boldsymbol{\mu}}}_{MM}-\boldsymbol{\mu} \right \|_2^2 \right ] = E\left [ \sum_{j=1}^p ({{\boldsymbol{\hat{\mu}}}_{MM}}_j-\boldsymbol{\mu}_j)^2 \right] =\sum_{j=1}^p   \sigma^2_{{{\boldsymbol{\hat{\mu}}}_{MM}}_j}\nonumber
\end{equation}

\cite{bose1993dispersion}, \cite{bose1995estimating} and \cite{mottonen2010asymptotic} investigated the asymptotic distribution for the $L_1-$median, and they obtained the following result in presence of normality:
\begin{equation}\label{assymptoticdistmedian}
\boldsymbol{\hat{\mu}}_{MM} \sim  N_p \left( \boldsymbol{\mu}, \frac{1}{n} \hat{A}^{-1} \hat{B} \hat{A}^{-1} \right),
\end{equation}

\noindent
where $\hat{A}(\mathbf{x}_{i})=\frac{1}{||\mathbf{x}_{i}||_2} \left( I_p - \frac{\mathbf{x}_{i} \mathbf{x}_{i}^T}{||\mathbf{x}_{i}||^2_2} \right)$ and $\hat{B}(\mathbf{x}_{i})=\frac{\mathbf{x}_{i} \mathbf{x}_{i}^T}{||\mathbf{x}_{i}||_2^2}$, with $\mathbf{x}_{i} \in \mathbb{R}^p$, for each $i=1,...,n$.

The numerator in the expression (\ref{alphaprop2}) can be obtained with the above property:
\begin{equation}\label{tracemuMM}
 E\left [ \left \| {\hat{\boldsymbol{\mu}}}_{MM}-\boldsymbol{\mu} \right \|_2^2 \right ] =trace \left( \frac{1}{n} \hat{A}^{-1} \hat{B} \hat{A}^{-1} \right)
\end{equation}

\subsection{Dispersion parameter}

Based on the median concept, which is a robust measure of location, one can define a robust measure of dispersion for a random variable $X$, which is the \textit{Median Absolute Deviation (MAD)} from the data's median:
\begin{equation}\label{MAD}
MAD(X)=median(|X-median(X)|),
\end{equation}

\cite{falk1997mad} showed the following relation, assuming normality, between the $MAD$ and the standard deviation $\sigma_X$:
\begin{equation}\label{eq:mad_sigma}
MAD(X)=\sigma_X \Phi ^{-1}(3/4),
\end{equation}

\noindent
where $\Phi$ denotes the standard normal cdf. Taking the square in (\ref{eq:mad_sigma}) we obtain a relation between the variance $\sigma_X^2$ and  $MAD^2(X)$:
\begin{equation}\label{eq:mad2_sigma2}
\sigma_X^2 = 2.198 \cdot MAD^2(X)
\end{equation}

Extending the idea of the $MAD$, a robust measure of dependence between two random variables $X$ and $Y$ is the \textit{comedian} (\cite{falk1997mad}):
\begin{equation}\label{comedian}
COM(X,Y)=med((X-med(X))(Y-med(Y)))
\end{equation}

The comedian generalizes the MAD, because $COM(X,X)=MAD^2(X)$, and also has the highest possible breakdown point (\cite{falk1997mad}). An important fact is that the comedian parallels the covariance, but the latter requires the existence of the first two moments of the two random variables, whereas the comedian always exists. Other known properties of the comedian are that it is symmetric, location invariant and scale equivariant. Furthermore, \cite{hall1985limit}  discussed about the strong consistency and asymptotic normality of the MAD, and \cite{falk1997mad} established similar results for the comedian.
 
Finally, a comedian matrix can be defined based on a multivariate version of (\ref{comedian}). Let $\textbf{x}=\{\mathbf{x}_{\cdot1},...,\mathbf{x}_{\cdot p}\}$ be the $n \times p$ data matrix with $n$ being the sample size and $p$ the number of variables. Then the comedian matrix  is defined as:
\begin{equation}
COM(\mathbf{x})=( \ COM(\mathbf{x}_{\cdot j},\mathbf{x}_{\cdot t}) \ ) \ j,t=1,...,p
\end{equation}

Note that from relation described in (\ref{eq:mad2_sigma2}), one can consider the adjusted comedian:
\begin{equation}\label{S^*}
\hat{S}_{CCM}=2.198 \cdot COM(\mathbf{x})
\end{equation}

Note that $\hat{S}_{CCM}$ is a robust alternative for the covariance matrix, but in general it is not positive (semi-) definite (see \cite{falk1997mad}). Since we need this property for inverting the covariance matrix in a Mahalanobis distance, we propose a shrinkage over $\hat{S}_{CCM}$, because of its advantage of providing always a positive definite and well-conditioned matrix.
Therefore, if a shrinkage estimator is considered in (\ref{shrinkgeneral}) for the dispersion parameter:
\begin{equation}\label{shrinkcov}
\hat{\Sigma}_{Sh}=(1-\eta) \hat{E} + \eta \hat{T},
\end{equation}

\noindent
we propose to use in (\ref{shrinkcov}), the estimator $\hat{E}=\hat{S}_{CCM}$.

Recall the previous sub-section \ref{subsectionlocation} in which we needed to provide a robust estimator for $\sigma_{\mathbf{x}_{\cdot j}}^2$, for each $j=1,...,p$ (Equation \ref{ExpectationmuCCM}) note that, because of the relation in (\ref{eq:mad2_sigma2}):
\begin{align}\label{sigmaCCM}
trace(\hat{S}_{CCM})&=\sum_{j=1}^p 2.198 \cdot COM(\mathbf{x}_{\cdot j},\mathbf{x}_{\cdot j})=\sum_{j=1}^p 2.198 \cdot MAD^2(\mathbf{x}_{\cdot j})=\sum_{j=1}^p \sigma^2_{\mathbf{x}_{\cdot j}}
\end{align}

Thus, when considering a shrinkage estimator of the component-wise median, in order to estimate the variance of $\boldsymbol{\hat{\mu}}_{CCM}$ needed in the expression (\ref{alphaprop1}) for the shrinkage intensity $\eta$, and according to the relation (\ref{ExpectationmuCCM}), we propose to estimate $\sum_{j=1}^p \sigma^2_{\mathbf{x}_{\cdot j}}$ using the $trace(\hat{S}_{CCM})$.

About the shrinkage target $\hat{T}$, several choices have been proposed in the literature. For example, \cite{ledoit2003improved} proposed a weighted average of the sample covariance matrix and a single-index covariance matrix. \cite{ledoit2003honey} proposed selecting the shrinkage target as a ``constant correlation matrix'', whose correlations are set equal to the average of all sample correlations. Finally, \cite{ledoit2004well} proposed to use a multiple of the identity matrix as the shrinkage target. The authors proved that the resulting shrinkage covariance matrix is well-conditioned, even if the sample covariance matrix is not. There is also another approach introduced by \cite{demiguel2013size}. The authors proposed a shrinkage estimator both for the covariance matrix and its inverse. The estimators were constructed as a convex combination of the sample covariance matrix or its inverse, respectively, and a scaled shrinkage target, which they consider the scaled identity matrix as \cite{ledoit2004well}. Therefore, we propose to use as shrinkage target $\hat{T}=\nu_{\Sigma}I$.
Thus (\ref{shrinkcov}) results in:
\begin{equation}\label{shrinkS*}
\hat{\Sigma}_{Sh(CCM)} = (1-\eta) \hat{S}_{CCM} + \eta \nu_{\Sigma} I
\end{equation}

Lastingly, the scaling parameter $\nu_{\Sigma}$ and the shrinkage intensity parameter $\eta$ in (\ref{shrinkS*}) need to be estimated. They both are chosen to minimize the expected quadratic loss as in \cite{ledoit2004well}:
\begin{equation}\label{eq:minsquarederror}
\begin{matrix}
\textup{min}_{\nu_{\Sigma},\eta} & E\left [ \left \| \hat{\Sigma}_{Sh} - \Sigma  \right \| ^2 \right ]\\ 
\textup{s.t.} & \hat{\Sigma}_{Sh} = (1-\eta) \hat{S}_{CCM} + \eta \nu_{\Sigma} I,
\end{matrix}
\end{equation}
\noindent
where $\left \|A\right \|^2=trace(AA^T)/p$. 

\begin{theorem}\label{prop3teo}
The solution of the problem (\ref{eq:minsquarederror}) is:
\begin{align*}
\nu_{\Sigma}&= trace(\hat{S}_{CCM})/p , \quad
\eta= \frac{E\left [ \left \| \hat{S}_{CCM}-\Sigma \right \|^2 \right ]}{E\left [ \left \| \hat{S}_{CCM}- \nu_{\Sigma}I \right \|^2 \right ]  }
\end{align*}
\end{theorem}

The proof can be found in Section 1 from the Supplementary Material. In practice, we propose to estimate the numerator of the expression for $\eta$ as \cite{ledoit2003honey}, \cite{ledoit2003improved} and \cite{ledoit2004well}, but considering $\hat{S}_{CCM}$ instead of the sample covariance matrix, as the estimator of $\Sigma$.

Note that the comedian matrix depends on centered data considering the component-wise median $\boldsymbol{\hat{\mu}}_{CCM}$. A special case of comedian matrix can be defined if the data are centered using a different location estimator. We propose to center the data using the other location estimators described in Subsection \ref{subsectionlocation}, i.e. the multivariate $L_1-$median $\boldsymbol{\hat{\mu}}_{MM}$, and the shrinkage estimators $\boldsymbol{\hat{\mu}}_{Sh(CCM)}$ and $\boldsymbol{\hat{\mu}}_{Sh(MM)}$. We will consider shrinkages over those special comedian matrices. 

\begin{enumerate}
\item $\hat{\Sigma}_{Sh(MM)}=(1- \eta) \hat{S}_{MM} + \eta \nu_{\Sigma} I$, with for $j,t=1,...,p$:

$\hat{S}_{MM}=2.198 \cdot COM_{MM}(\mathbf{x})=2.198\cdot (med((\mathbf{x}_{\cdot j} - (\boldsymbol{\hat{\mu}}_{MM})_j ) (\mathbf{x}_{\cdot t} - (\boldsymbol{\hat{\mu}}_{MM})_t ))$

\item $\hat{\Sigma}_{Sh(Sh(CCM))} = (1-\eta) \hat{S}_{Sh(CCM)} + \eta \nu_{\Sigma} I$, with for $j,t=1,...,p$:

\noindent
$\hat{S}_{Sh(CCM)}=2.198\cdot COM_{Sh(CCM)}(\mathbf{x})=2.198\cdot (med((\mathbf{x}_{\cdot j} - (\boldsymbol{\hat{\mu}}_{Sh(CCM)})_j  ) (\mathbf{x}_{\cdot t} - (\boldsymbol{\hat{\mu}}_{Sh(CCM)})_t  ))$
 
\item $\hat{\Sigma}_{Sh(Sh(MM))}=(1- \eta) \hat{S}_{Sh(MM)} + \eta \nu_{\Sigma} I$, with for $j,t=1,...,p$:

\noindent
$\hat{S}_{Sh(MM)}=2.198\cdot COM_{Sh(MM)}(\mathbf{x})=2.198\cdot (med((\mathbf{x}_{\cdot j} - (\boldsymbol{\hat{\mu}}_{Sh(MM)})_j ) (\mathbf{x}_{\cdot t} - (\boldsymbol{\hat{\mu}}_{Sh(MM)})_t ))$
\end{enumerate}

The optimal expression for the parameters $\eta$ and $\nu_{\Sigma}$ in the above cases is analogous to the Proposition \ref{prop3teo}, but considering in each case the sample estimator as the corresponding special comedian matrix.

\subsection{Proposed Robust Mahalanobis Distances}

A robust Mahalanobis distance can be defined as in (\ref{eq:robustmaha}), for each of the following 6 possible combinations for the location and the dispersion estimators (see Table \ref{tab:combinations}). Note that they are meaningful combinations because the shrinkage estimator of dispersion is made upon a special comedian matrix closely based on the location estimator jointly considered for defining the \textit{RMD}.

\begin{table}[htbp]

  \caption{Combinations of location and dispersion}
    \begin{tabular}{l|l|l|l|l|l|l}
    Name  & \multicolumn{1}{l}{RMDv1} & \multicolumn{1}{l}{RMDv2} & \multicolumn{1}{l}{RMDv3} & \multicolumn{1}{l}{RMDv4} & \multicolumn{1}{l}{RMDv5} & \multicolumn{1}{l}{RMDv6} \\
    \midrule
    $\hat{\mu}_R$  &  $\hat{\mu}_{CCM}$      &  $\hat{\mu}_{Sh(CCM)}$      & $\hat{\mu}_{Sh(CCM)}$      &  $\hat{\mu}_{MM}$      &  $\hat{\mu}_{Sh(MM)}$      & $\hat{\mu}_{Sh(MM)}$   \\
    $\hat{\Sigma}_R$  &  $\hat{\Sigma}_{Sh(CCM)}$     &       $\hat{\Sigma}_{Sh(CCM)}$  &   $\hat{\Sigma}_{Sh(Sh(CCM))}$      &  $\hat{\Sigma}_{Sh(MM)}$     &       $\hat{\Sigma}_{Sh(MM)}$  &   $\hat{\Sigma}_{Sh(Sh(MM))}$       \\
    \end{tabular}%
  \label{tab:combinations}%
\end{table}%

For all our proposed combinations, the threshold considered to detect the outliers is the $\chi^2_{p;0.975}$ quantile.

\section{Simulation results}
\subsection{Normal distribution}\label{simusNormal}

A simulation study is performed considering a $p-$dimensional random variable $X$ following a contaminated multivariate normal distribution given as a mixture of normals of the form $(1-\alpha)N(\mathbf{0},I)+ \alpha N(\delta  \mathbf{e} , \lambda I)$, where $\mathbf{e}$ denotes the $p-$dimensional vector of ones. This model is analogous to the one used by \cite{rousseeuw1999fast}, \cite{pena2001multivariate}, \cite{filzmoser2005multivariate}, \cite{pena2007combining}, \cite{maronna2002robust} and \cite{sajesh2012outlier}. This experiment has been conducted for different values of the sample-space dimension $p=5,10,30$, and the chosen sample size in relation to the dimension was $n=100,100,500$, respectively. The contamination levels were $\alpha=0,0.1,0.2,0.3$, the distance of the outliers $\delta=5,10$ and the concentration of the contamination $\lambda=0.1,1$. For each set of values, 100 random sample repetitions have been generated.

For the methods mentioned in previous sections some measures are studied: the correct detection rates (c) and the false detection rates (f). The method \textit{MCD}  refers to the \textit{RMD} based on the \textit{MCD} estimator and with the classical threshold, the method \textit{Adj MCD} refers to the latter distance considering the adjusted quantile of \cite{filzmoser2005multivariate}, the method \textit{Kurtosis} refers to the \cite{pena2007combining} approach, the method \textit{OGK} refers to the Orthogonalized Gnanadesikan-Kettenring method proposed by \cite{maronna2002robust} and \textit{COM} is the Comedian method proposed by \cite{sajesh2012outlier}. We have also presented the results for the collection \textit{RMDv1}-\textit{RMDv6} proposed in Table \ref{tab:combinations}. All simulations were performed in Matlab. Section 2 in the Supplementary Material shows the tables corresponding to all simulation scenarios. Here we show only the most significant and representative results.
\begin{table}[htbp]
  \centering
  \caption{Correct detection rates, with Normal distribution.}
   \resizebox{15cm}{!} {
    \begin{tabular}{c|cccccccccccc}
    $\delta=5$ &   &  $\lambda=0.1$     &       &       &       &       &       &       &       &       &       &  \\
    \midrule
    $p$     & $\alpha$ & MCD   & Adj MCD & Kurtosis & OGK   & COM   & RMDv1 & RMDv2 & RMDv3 & RMDv4 & RMDv5 & RMDv6 \\
    \midrule
    5     & 0.1   & 1     & 1     & 0,9000 & 1     & 1     & 1     & 1     & 1     & 1     & 1     & 1 \\
          & 0.2   & 0,8700 & 0,8700 & 0,5100 & 0,9500 & 0,9941 & 1     & 1     & 1     & 1     & 1     & 1 \\
          & 0.3   & 0,0600 & 0,0600 & 0,9800 & 0,1500 & 0,5719 & 0,8766 & 0,8782 & 0,8782 & 0,9146 & 0,9090 & 0,9130 \\
    \midrule
    10    & 0.1   & 0,9900 & 0,9900 & 0,8600 & 1     & 1     & 1     & 1     & 1     & 1     & 1     & 1 \\
          & 0.2   & 0,2800 & 0,2800 & 0,4600 & 0,9416 & 1     & 1     & 1     & 1     & 1     & 1     & 1 \\
          & 0.3   & 0     & 0     & 0,9900 & 0,1612 & 0,7205 & 0,8774 & 0,8747 & 0,8750 & 0,9711 & 0,9672 & 0,9711 \\
    \midrule
    30    & 0.1   & 0,1900 & 0,1900 & 1     & 1     & 1     & 1     & 1     & 1     & 1     & 1     & 1 \\
          & 0.2   & 0     & 0     & 0,1   & 1     & 1     & 1     & 1     & 1     & 1     & 1     & 1 \\
          & 0.3   & 0     & 0     & 0,6100 & 0,0100 & 0,9407 & 0,5308 & 0,5275 & 0,5286 & 0,9990 & 0,9988 & 0,9991 \\
    \end{tabular}%
    }
    \label{tab:c_normallambda5}%
\end{table}%
Nevertheless, the tables show general outcomes. For example, \textit{Adj MCD}, actually improves \textit{MCD} with respect to the ``f'', lowering it, and in most cases  maintaining the same ``c''. Although, in other cases it also slightly lowers the ``c''. On the other hand, the false detection rates in case of no contamination are sufficiently low for all methods, but our proposed collection shows the lowest values especially in high dimension, actually here the best performance is observed for \textit{RMDv6}. With certain percent of contamination, the worst behavior of our proposed methods is when dimension is low and the highest percentage of outliers are considered to be near the center of the data. This matter can be seen in Table \ref{tab:c_normallambda5} which corresponds to the correct detection rates. In this case \textit{Kurtosis} has better performance. Although, this only happens in two cases, and in all other cases \textit{MCD}, \textit{Adj MCD}, \textit{Kurtosis} and \textit{OGK} are the ones with the worst behavior. Meanwhile, \textit{COM} is a good competitor, but the best overall performance is made by \textit{RMDv6} especially in high dimension.

Another situation is when outliers are far from the center of the data, i.e. $\delta=10$. This scenario is shown in Table \ref{tab:c_normallambda10}. It is clear that our proposed methods lead to the best performance, achieving $100\%$ of correct detection rate, for all dimension and percentage of contamination considered. Other tables about the ``c'' can be found in the Supplementary Material, as well as the false detection rate tables, which show that in the vast majority of cases our proposal have an ``f'' value equal to zero and when not, a value very close to zero, which is what is desirable.

\begin{table}[htbp]
  \centering
  \caption{Correct detection rates, with Normal distribution.}
   \resizebox{15cm}{!} {
    \begin{tabular}{c|cccccccccccc}
   $\delta=10$ &  $\alpha$ &    $\lambda=1$   &       &       &       &       &       &       &       &       &       &  \\
   \midrule
    $p$     & $\alpha$ & MCD   & Adj MCD & Kurtosis & OGK   & COM   & RMDv1 & RMDv2 & RMDv3 & RMDv4 & RMDv5 & RMDv6 \\
    \midrule
    5     & 0.1   & 1     & 1     & 1     & 1     & 1     & 1     & 1     & 1     & 1     & 1     & 1 \\
          & 0.2   & 0,8480 & 0,8465 & 0,9900 & 1     & 1     & 1     & 1     & 1     & 1     & 1     & 1 \\
          & 0.3   & 0,2190 & 0,1976 & 0,9307 & 0,9591 & 0,9991 & 1     & 1     & 1     & 1     & 1     & 1 \\
    \midrule
    10    & 0.1   & 1     & 1     & 0,9800 & 1     & 1     & 1     & 1     & 1     & 1     & 1     & 1 \\
          & 0.2   & 0,8623 & 0,8548 & 0,6558 & 1     & 1     & 1     & 1     & 1     & 1     & 1     & 1 \\
          & 0.3   & 0,2280 & 0,2046 & 0,4618 & 0,9911 & 1     & 1     & 1     & 1     & 1     & 1     & 1 \\
    \midrule
    30    & 0.1   & 1     & 1     & 0,8919 & 1     & 1     & 1     & 1     & 1     & 1     & 1     & 1 \\
          & 0.2   & 0,4879 & 0,4654 & 0,0125 & 1     & 1     & 1     & 1     & 1     & 1     & 1     & 1 \\
          & 0.3   & 0,0810 & 0,0509 & 0,1087 & 1     & 1     & 1     & 1     & 1     & 1     & 1     & 1 \\
    \end{tabular}%
    }
    \label{tab:c_normallambda10}%
\end{table}%

\subsection{$t_3$-distribution}\label{simusT}

In order to check the behavior of the methods when the distribution deviates from normality, a simulation study is performed considering a $p-$dimensional random variable $X$ following a contaminated multivariate $t$-distribution with $3$ degrees of freedom of the form $(1-\alpha)T_3(\mathbf{0},I)+ \alpha T_3(\delta  \mathbf{e} , \lambda I)$. The first parameter of the notation of $T_3(\cdot,\cdot)$ refers to the mean and the second one to the covariance matrix. The parameters for the contamination are the same considered above and the same measures ``c'' and ``f'' are studied. All the results can be founded in the tables from Section 2 in the Supplementary Material. It should be noted the unsatisfactory behavior of our competitors with respect to the ``c'' especially in high dimension or with large contamination level, meanwhile in most cases we attain a $100\%$ correct detection rate. With respect to the ``f'' value, all methods show non-zero ``f'' values, and the best performance is showed by \textit{COM} and our proposed methods.

\subsection{Exponential distribution}\label{simusE}

We considered also a $p-$dimensional random variable $X$ following a contaminated multivariate exponential distribution given as a mixture $(1-\alpha)Exp(\mathbf{0})+ \alpha Exp(\delta  \mathbf{e})$. The parameter of the notation $Exp(\cdot)$ refers to the mean. This case is analogous to the previous ones, with the difference that only the schemes associated with the distance of the outliers are considered. The tables with the results can be found in Section 2 in the Supplementary Material, and it can be seen that our methods achieve $100\%$ of correct detection rate in the majority of cases, except in some cases when dimension is low. Actually in all cases the best performance is showed by \textit{RMDv6}. On the other hand, all methods show more or less the same ``f'' value.

\subsection{Summary and selection of one of our proposed distances}

In the simulation study, for each contamination scheme we have also calculated a measure called F-score (\cite{goutte2005probabilistic}, \cite{sokolova2006beyond}, \cite{powers2011evaluation}), often used in Engineering, which is a measure of a test's accuracy. Its expression is F-score$ =2PR/(P+R)$, where $P$ is called precision  and $R$ is known as the recall. The precision $P$ is the number of correct detected outliers divided by the total number of detected outliers, and the recall $R$ is the number of correct detected outliers divided by the real total number of outliers. Thus, this measure provides a trade-off between the two desired outcomes: a high rate of correctly identified outliers and a low rate of observations mislabel as outliers. The results are not included in the paper for avoiding large extension, but the method with the overall classification between the top 3 best positions ranking with respect to the F-score, is method \textit{RMDv6}.

It is clear the out-performance of our proposed methods with gaussian data, especially in high dimension and even when we deviate from the normality assumption, for example when considering skewed and heavy-tailed distributions like the multivariate $t_3$-distribution and the multivariate exponential distribution. From all of our six proposed robust distances, the one that shows the best results in the vast majority of cases is \textit{RMDv6}. Thus, we decided to select it as the best one in the matter of performance, and from now on we will refer to it as \textit{RMD-Shrinkage}. Now we will proceed to study some properties like the behavior under correlated data, the affine equivariance, the breakdown value, and the computational time.

\subsection{Correlation and affine equivariance}

Consider $X=\lbrace \mathbf{x_1},...,\mathbf{x_n}\rbrace$ and a pair of multivariate location and covariance estimators $(m,S)$. In general, these estimators are called affine equivariant if for any nonsingular matrix $A$ it holds that:
\begin{align}\label{affeq}
m_A=m(X_A)=Am(X), \quad  S_A=S(X_A)=S(X)A^T
\end{align}
The affine transformation of $X$ is $X_A=\lbrace A\mathbf{x_1},...,A\mathbf{x_n}\rbrace$. Affine equivariance implies that the estimator transforms well under any nonsingular reparametrization of the space of the $\mathbf{x}_i$. The data might for instance be rotated, translated or rescaled (for example through a change of the measurement units).

The method \textit{RMD-Shrinkage} is ultimately based on not affine equivariant estimators which are the $L_1$-median (\cite{lopuhaa1991breakdown}) and the comedian matrix. However, the $L_1$-median is orthogonal equivariant, i.e. it satisfies Equation (\ref{affeq}) with $A$ any orthogonal matrix $(A'=A^{-1})$. This implies that the $L_1$-median transforms appropriately under all transformations that preserve Euclidean distances (such as translations, rotations and reflections). About the comedian matrix, which always exists, it is symmetric, location invariant and scale equivariant (\cite{falk1997mad}), i.e. $COM(X, \mathbf{a}Y+\mathbf{b})=\mathbf{a} COM(X,Y)=\mathbf{a} COM(Y,X)$. Since the proposed method is not affine equivariant, it is important to investigate the behavior under correlated data. \cite{devlin1981robust} used a correlation matrix $P$ for generating Monte Carlo data from different distributions of moderate dimension $p=6$. The matrix has the form:
\begin{equation}
P=\begin{bmatrix}
P_1 & 0\\ 
0 & P_2
\end{bmatrix}
\end{equation}
\begin{equation}
P_1=\begin{bmatrix}
1 & 0.95 & 0.3\\ 
0.95 & 1 & 0.1\\
0.3 & 0.1 & 1
\end{bmatrix}
, \quad P_2=\begin{bmatrix}
1 & -0.499 & -0.499\\ 
-0.499& 1 & -0.499\\
-0.499 & -0.499 & 1
\end{bmatrix}
\end{equation}
The reason for the selection of the matrix $P$ is because the dimension is large enough to study multivariate estimators and the range of correlation values is large. This way the differences in the abilities of the methods to detect outliers from highly correlated data can be observed. For the simulations, $n=100$ observations were generated from a mixture of Normals $(1-\alpha)N(\mathbf{0},P)+\alpha N(5\mathbf{e},P)$. The contamination level $\alpha=10\%,20\%,30\%$.

Table \ref{tab:corr} shows that the correct and false detection rates of \textit{MCD}, \textit{Adj MCD}, \textit{Kurtosis} and \textit{OGK} are worse than that of our proposal . On the other hand,  \textit{COM} show more or less the same behavior in case of $10\%$ and $20\%$ of contamination, and slightly worse than our proposal when the contamination level increases to $30\%$. The methods \textit{MCD}, \textit{Adj MCD} and \textit{Kurtosis} are affine equivariant, while \textit{OGK} and \textit{COM} are not. Hence, the proposed procedure \textit{RMD-Shrinkage} is more efficient than other affine and not affine equivariant methods in case of correlated datasets. Also the false detection rate is very low even in this case of presence of correlation.

\begin{table}[htbp]
  \centering
  \caption{Simulation results for correlated data.}
  \resizebox{15cm}{!} {
    \begin{tabular}{c|cc|cc|cc|cc|cc|cc}
          & MCD   &       & Adj MCD &       & Kurtosis &       & OGK   &       & COM   &       & RMD-Shrinkage &  \\
    \midrule
    $\alpha$ & c     & f     & c     & f     & c     & f     & c     & f     & c     & f     & c     & f \\
    \midrule
    0.1   & 1     & 0,0397 & 1     & 0,0226 & 1     & 0,0371 & 1     & 0,0736 & 1     & 0,0025 & 1     & 0,0128 \\
    0.2   & 0,8659 & 0,0127 & 0,8565 & 0,0062 & 0,8771 & 0,0453 & 0,9792 & 0,0533 & 1     & 0,0011 & 1     & 0,0013 \\
    0.3   & 0,1504 & 0,0762 & 0,1238 & 0,0614 & 0,8186 & 0,0443 & 0,4780 & 0,0460 & 0,8302 & 0,0001 & 0,9274 & 0 \\
    \end{tabular}%
    }
  \label{tab:corr}%
\end{table}%

Affine equivariance of the estimators is equivalent to say that the robust Mahalanobis distance is affine invariant:
$$RMD(\mathbf{Ax_i},\mathbf{m_A})=(\mathbf{Ax_i}-\mathbf{m_A})^T \mathbf{S_A}^{-1} (\mathbf{Ax_i} - \mathbf{m_A})=RMD(\mathbf{x_i},\mathbf{m})$$

\cite{maronna2002robust} and \cite{sajesh2012outlier} proposed to investigate the lack of equivariance with transformed data, by simulations. We study the same for our proposal. They propose to generate random matrices as $A=TD$, where $T$ is a random orthogonal matrix and $D=diag(u_1,...,u_p)$, where the $u_j$'s are independent and uniformly distributed in $(0,1)$. Then, the proposed simulations consist on affinely transform each generated data matrix $X$ in each repetition, by applying the random matrix of transformation $A$ to $X$, in order to obtain $X_A$.

The contamination scheme consist in a mixture of normals $(1-\alpha)N(\mathbf{0},I)+\alpha N(\delta \mathbf{e}, \lambda I)$. The dimension $p=5,10,30$, with sample size $n=100,100,500$ respectively, the contamination level $\alpha=10\%,20\%,30\%$, the distance of the outliers $\delta=5,10$, and the concentration of the contamination $\lambda=0.1,1$. Table \ref{tab:1} shows the obtained results about the correct and false detection rates.

\begin{table}[htbp]
  \centering
  \caption{Correct and false detection rates of RMD-Shrinkage for transformed data.}
   \resizebox{9cm}{!} {
    \begin{tabular}{cccrrrr}
          &       &       & \multicolumn{1}{c}{$\delta=5$} &       & \multicolumn{1}{c}{$\delta=10$} &  \\
\cmidrule{4-7}          & $p$     & $\alpha$ & \multicolumn{1}{c}{c} & \multicolumn{1}{c}{f} & \multicolumn{1}{c}{c} & \multicolumn{1}{c}{f} \\
    \midrule
    $\lambda=0.1$ & 5     & 0.1   & 1 & 0,0454 & 1 & 0,0455 \\
          &       & 0.2   & 1 & 0,0155 & 1 & 0,0165 \\
          &       & 0.3   & \textbf{0,9709} & 0,0034 & 1 & 0,0023 \\
          & 10    & 0.1   & 1 & 0,0328 & 1 & 0,0252 \\
          &       & 0.2   & 1 & 0,0088 & 1 & 0,0062 \\
          &       & 0.3   & \textbf{0,9844} & 0,0023 & 1 & 0,0009 \\
          & 30    & 0.1   & 1 & 0,0089 & 1 & 0,0074 \\
          &       & 0.2   & 1 & 0,0006 & 1 & 0,0003 \\
          &       & 0.3   & 1 & 0 & 1 & 0 \\
    \midrule
    $\lambda=1$ & 5     & 0.1   & 1 & 0,0451 & 1 & 0,0400 \\
          &       & 0.2   & 1 & 0,0189 & 1 & 0,0120 \\
          &       & 0.3   & \textbf{0,9344} & 0,0039 & 1 & 0,0046 \\
          & 10    & 0.1   & 1 & 0,0282 & 1 & 0,0279 \\
          &       & 0.2   & 1 & 0,0113 & 1 & 0,0071 \\
          &       & 0.3   & \textbf{0,9872} & 0,0020 & 1 & 0,0010 \\
          & 30    & 0.1   & 1 & 0,0093 & 1 & 0,0072 \\
          &       & 0.2   & 1 & 0,0006 & 1 & 0,0004 \\
          &       & 0.3   & 1 & 0 & 1 & 0 \\
    \end{tabular}%
    }
  \label{tab:1}%
\end{table}%

 As it can be observed, even under affine transformations, \textit{RMD-Shrinkage} is able to detect all the outliers, except for a few cases (in bold type) that corresponds to large contamination level ($30\%$) in case of outliers close to the center of the distribution. However, it can be noted that these cases improve in performance when dimension increases.

\subsection{Breakdown value}

The maximum proportion of outliers that the estimator can safely tolerate is known as the breakdown value. For an outlier detection method, it can be defined as the maximum $\alpha^*$ outliers that the procedure can successfully detect, so that if $\alpha>\alpha^*$ the method will fail to identify most of the true outliers and it will falsely detect many inliers, reducing drastically the``c'' and inflating the ``f''. Thus, it is necessary to use the correct and false detection rates for studying the breakdown value of the outlier detection procedure.
\begin{table}[htbp]
  \centering
  \caption{Simulation results for breakdown value.}
  \resizebox{9cm}{!} {
    \begin{tabular}{|cc|cc|cc|}
    \toprule
    $n=1000$ & \multicolumn{1}{c}{} & Symmetric & \multicolumn{1}{c}{} & Asymmetric &  \\
    \midrule
   $ p $    & \multicolumn{1}{c}{$\alpha$} & c     & \multicolumn{1}{c}{f} & c     & f \\
    \midrule
    10    & 0.1   & 1     & 0,0055 & 1     & 0,0047 \\
          & 0.2   & 1     & 0,0001 & 1     & 0,0002 \\
          & 0.3   & 1     & 0     & 1     & 0 \\
          & 0.4   & 1     & 0     & 1     & 0 \\
          & 0.45  & 1     & 0     & 1     & 0 \\
    \midrule
    30    & 0.1   & 1     & 0,0002 & 1     & 0,0002 \\
          & 0.2   & 1     & 0     & 1     & 0 \\
          & 0.3   & 1     & 0     & 1     & 0 \\
          & 0.4   & 1     & 0     & 1     & 0 \\
          & 0.45  & 1     & 0     & 1     & 0 \\
    \midrule
    50    & 0.1   & 1     & 0     & 1     & 0 \\
          & 0.2   & 1     & 0     & 1     & 0 \\
          & 0.3   & 1     & 0     & 1     & 0 \\
          & 0.4   & 1     & 0     & 1     & 0 \\
          & 0.45  & 1     & 0     & 1     & 0 \\
	\midrule    
    80    & 0.1   & 1     & 0     & 1     & 0 \\
          & 0.2   & 1     & 0     & 1     & 0 \\
          & 0.3   & 1     & 0     & 1     & 0 \\
          & 0.4   & 1     & 0     & 1     & 0 \\
          & 0.45  & 1     & 0     & 1     & 0 \\
    \midrule
    100   & 0.1   & 1     & 0     & 1     & 0 \\
          & 0.2   & 1     & 0     & 1     & 0 \\
          & 0.3   & 1     & 0     & 1     & 0 \\
          & 0.4   & 1     & 0     & 1     & 0 \\
          & 0.45  & 1     & 0     & 1     & 0 \\
    \bottomrule
    \end{tabular}%
    }
  \label{tab:2}%
\end{table}%

Analogously as \cite{sajesh2012outlier}, we consider two forms of contamination: $\alpha$ percent symmetric, for which the $i$th observation is multiplied by $100i$, and $\alpha$ percent asymmetric, for which the $i$th observation is replaced by $(100i)\mathbf{e}$, $i=1,...,n\alpha$, where $\mathbf{e}=(1,...,1)$. In the first case the outliers are symmetrically distributed, and asymmetrically in the second case.

The dimensions considered are $p=10,30,50,80,100$ and the sample size $n=1000$. The contamination level $\alpha=10\%,20\%,30\%,40\%,45\%$. Table \ref{tab:2} gives the resulting correct and false detection rates for both forms of contamination. For all the values of $p$ and $\alpha$ the method attains a $100\%$ correct detection rate.  The maximum false detection rate for symmetric case is $0.55\%$ and for the asymmetric case is $0.47\%$. These results, even for high values of $\alpha$, indicates that \textit{RMD-Shrinkage} can robustly detect large amount of contamination.

\newpage
\subsection{Computational times}

Table \ref{tab:times} show the resulting computational times in seconds for the Normal case when outliers are close to the center of the data and they are concentrated. The other tables can be founded in the Supplementary Material.

\begin{table}[htbp]
  \centering
  \caption{Computational times with Normal data.}
  \resizebox{12cm}{!} {
    \begin{tabular}{c|ccccccc}
    \multicolumn{1}{c}{$\delta=5$} & $\lambda=0.1$ &       &       &       &       &       &  \\
$ p  $   & $\alpha$ & MCD   & Adj MCD & Kurtosis & OGK   & COM   & RMD-Shrinkage \\
    \midrule
    5     & 0.1   & 1,0951 & 0,7670 & 0,1880 & 0,1087 & 0,0228 & 0,0096 \\
          & 0.2   & 0,7619 & 0,7910 & 0,0499 & 0,0203 & 0,0088 & 0,0085 \\
          & 0.3   & 0,7605 & 0,8304 & 0,0266 & 0,0196 & 0,0089 & 0,0074 \\
          & \textbf{Mean} & \textbf{0,8725} & \textbf{0,7961} & \textbf{0,0882} & \textbf{0,0495} & \textbf{0,0135} & \textbf{0,0085} \\
    \midrule
    10    & 0.1   & 1,3184 & 0,9970 & 0,2191 & 0,1626 & 0,0247 & 0,0200 \\
          & 0.2   & 1,0329 & 1,0477 & 0,1358 & 0,0793 & 0,0120 & 0,0118 \\
          & 0.3   & 0,9685 & 1,0641 & 0,0482 & 0,0865 & 0,0128 & 0,0108 \\
          & \textbf{Mean} & \textbf{1,1066} & \textbf{1,0363} & \textbf{0,1344} & \textbf{0,1095} & \textbf{0,0165} & \textbf{0,0142} \\
    \midrule
    30    & 0.1   & 6,2387 & 6,0934 & 0,7154 & 0,8969 & 0,2000 & 0,2206 \\
          & 0.2   & 5,8676 & 6,3999 & 1,4635 & 0,8158 & 0,1687 & 0,1804 \\
          & 0.3   & 5,9453 & 7,0405 & 1,6572 & 0,8407 & 0,1669 & 0,1674 \\
          & \textbf{Mean} & \textbf{6,0172} & \textbf{6,5113} & \textbf{1,2787} & \textbf{0,8511} & \textbf{0,1785} & \textbf{0,1895} \\
    \end{tabular}%
    }
  \label{tab:times}%
\end{table}%

On average, the fastest methods are \textit{COM} and \textit{RMD-Shrinkage} with very similar computational times. When compared to the \textit{MCD} and its adjusted version \textit{Adj MCD}, the latters are much more slower than our proposal. The \textit{Kurtosis} and \textit{OGK}  are not that slower but they show worse times than ours. Thus, \textit{RMD-Shrinkage} shows competitive computational times.

\section{Real dataset}\label{sectionreal}

The proposed \textit{RMD's} are applied to a real dataset to evaluate their performance. The following dataset was taken from the \textit{UCI Knowledge Discovery in Databases Archive} \citep{bay1999uci}. Specifically, we have chosen the \textit{Breast Cancer Wisconsin (Diagnostic) Data Set} (WDBC). Features are computed from a digitized image of a fine needle aspirate of a breast mass. They describe 30 characteristics of the cell nuclei present in the image, for 569 samples, from which 357 are benign and 212 malign. We propose to study only the 357 benign data, as \cite{maronna2002robust}. Therefore, this example has dimension $p=30$ and sample size $n=357$. We applied each method for detecting outliers and we retained the results, along with the computational times. 

\begin{figure}[H]
     \centering
     \includegraphics[width=0.8\textwidth]{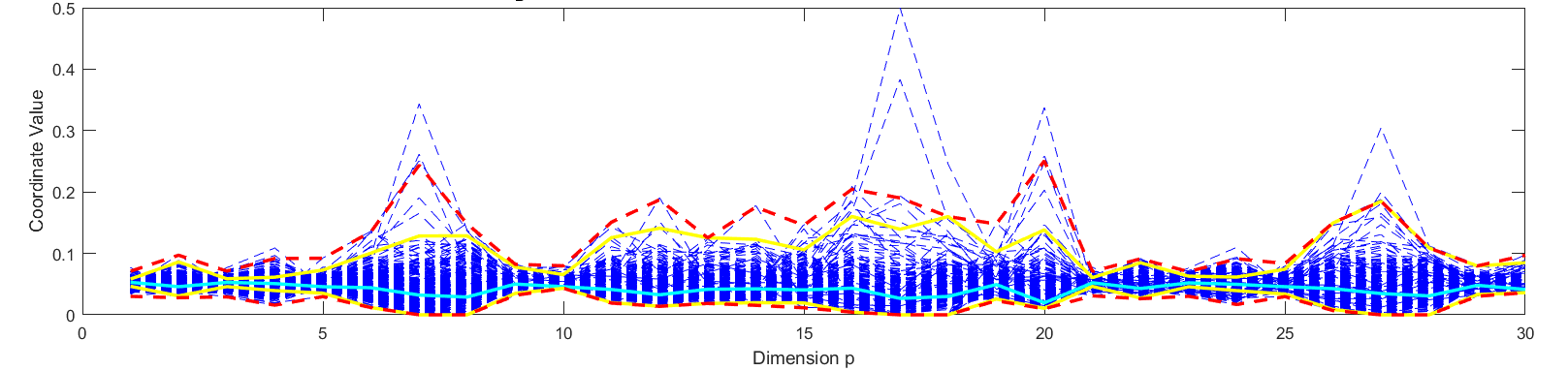}
     \caption{{\footnotesize Standardized data with the ``multivariate boxplot''.}}            
     \label{databoxplot}
\end{figure}

In order to interpret the outcome, we show the standardized data (after the detection) only for better visualization aim. We have also plotted the multivariate $L_1$ median and a kind of ``multivariate boxplot'', which is based on the idea from \cite{sun2011functional} method, but for finite dimensional. What the ``box'' would be is constructed sorting the data according to their $L_1$ depth value. The corresponding $Q_1$ and $Q_3$ ``quartiles'' delimiting the ``box'' are in fact the minimum and maximum values for each coordinate taking only into account the $50\%$ of the most central data. Thus, the ``fences'' can be constructed with the same approach $F_1=Q_1-1.5RI$ and $F_2=Q_3+1.5RI$, where the ``interquartile range'' is $RI=Q_3-Q_1$. Then, we can look for each method's result how many detected outliers are inside the ``fences'' for all their coordinates, and how many are outside the ``fences''. Figure \ref{databoxplot} shows the data in blue color plotted in parallel coordinates (\cite{inselberg1990parallel}, \cite{wegman1990hyperdimensional}, \cite{inselberg2009parallel}), the ``box'' delimiting the $50\%$ of most central data in yellow color, the ``fences'' in red  and the multivariate $L_1-$median in cyan.

\begin{table}[H]
\caption{Detected outliers.}
\begin{subtable}{1\textwidth}
	\caption{Inside and outside the fences.}
	\centering
	\resizebox{6.5cm}{!} {
    \begin{tabular}{|l|rrr|}
    \toprule
    \textbf{Method} & \multicolumn{1}{l}{\textbf{Out\_in}} & \multicolumn{1}{l}{\textbf{Out\_out}} & \multicolumn{1}{l|}{\textbf{Out\_total}} \\
    \midrule
    MCD   & 72    & 4     & 76 \\
    Adj MCD & 64    & 4     & 68 \\
    Kurtosis & 158   & 4     & 162 \\
    OGK & 144    & 4     & 148 \\
    COM & 59    & 4     & 63 \\
    RMD-Shrinkage & 25    & 3     & 28 \\
    \bottomrule
    \end{tabular}
    }
	\label{tab:fencesresults}
\end{subtable}
\bigskip
\begin{subtable}{1\textwidth}
	\caption{Inside the $50\%$ of the most central data.}
	\centering
	\resizebox{5.5cm}{!} {
    \begin{tabular}{|l|rr|}
    \toprule
    \textbf{Method} & \multicolumn{1}{l}{\textbf{Out\_in}}  & \multicolumn{1}{l|}{\textbf{Out\_total}} \\
    \midrule
    MCD   & 29   & 76 \\
    Adj MCD & 27     & 68 \\
    Kurtosis & 65      & 162 \\
    OGK & 58        & 148 \\
    COM & 20       & 63 \\
    RMD-Shrinkage & 7       & 28 \\
    \bottomrule
    \end{tabular}
    }
    \label{tab:qresults}
\end{subtable}
\end{table}%

Table \ref{tab:fencesresults} shows the detected outliers by each method. Outside the ``fences'' there are 3 or 4 for all the methods. Also, the method \textit{Kurtosis} detected 162 outliers out of the 357 data. More or less like \textit{OGK}, which detected 148. Furthermore, our method \textit{RMD-Shrinkage} is the one that label less amount of data as outliers. Table \ref{tab:qresults} shows how many outliers belong to the $50\%$ of the most central data, according to the $L_1$-median. We can investigate the shape of the detected outliers inside the ``multivariate box'', in order to see if they are similar or near to the median, or if they have a distinct shape.
Figure \ref{outliersrealdata} shows the shape of some of our competitors detected outliers that belong to the $50\%$ of the most central data. In cyan color is the multivariate median, in yellow color the ``box'' and in blue color the detected outlier. For all of our competitors there seems to be some outliers having a shape very similar to the multivariate median or close to it for all the values of its components, leading us to think that maybe they are detecting too much. However, in the figure associated with \textit{RMD-Shrinkage}, we can see that all outliers are quite different than the multivariate median, in fact, they might be shape outliers. For a final argument, we can say that all the outliers that belong to the ``multivariate box'' which are detected by the method \textit{RMD-Shrinkage}, are actually detected by our competitors (the intersection matches), so this also makes us think that our method detects just enough.

\begin{figure}[H]
     \centering
     \includegraphics[width=0.6\textwidth]{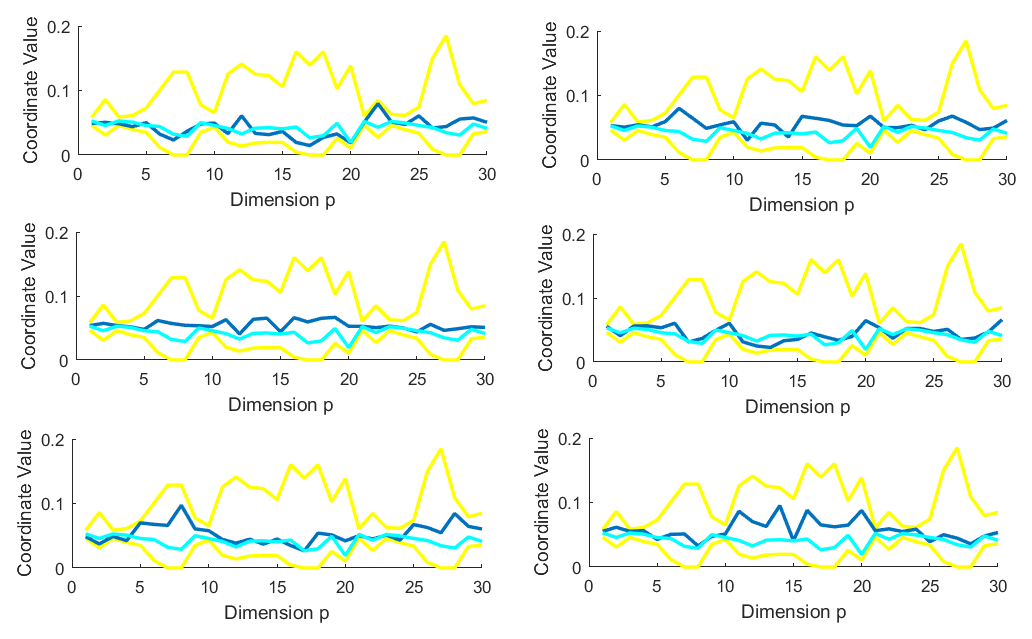}
     \caption{{\footnotesize Some of our competitors detected outliers belonging to the $50\%$ of the most central
data.}}            
     \label{outliersrealdata}
\end{figure}

\begin{figure}[H]
     \centering
     \includegraphics[width=0.8\textwidth]{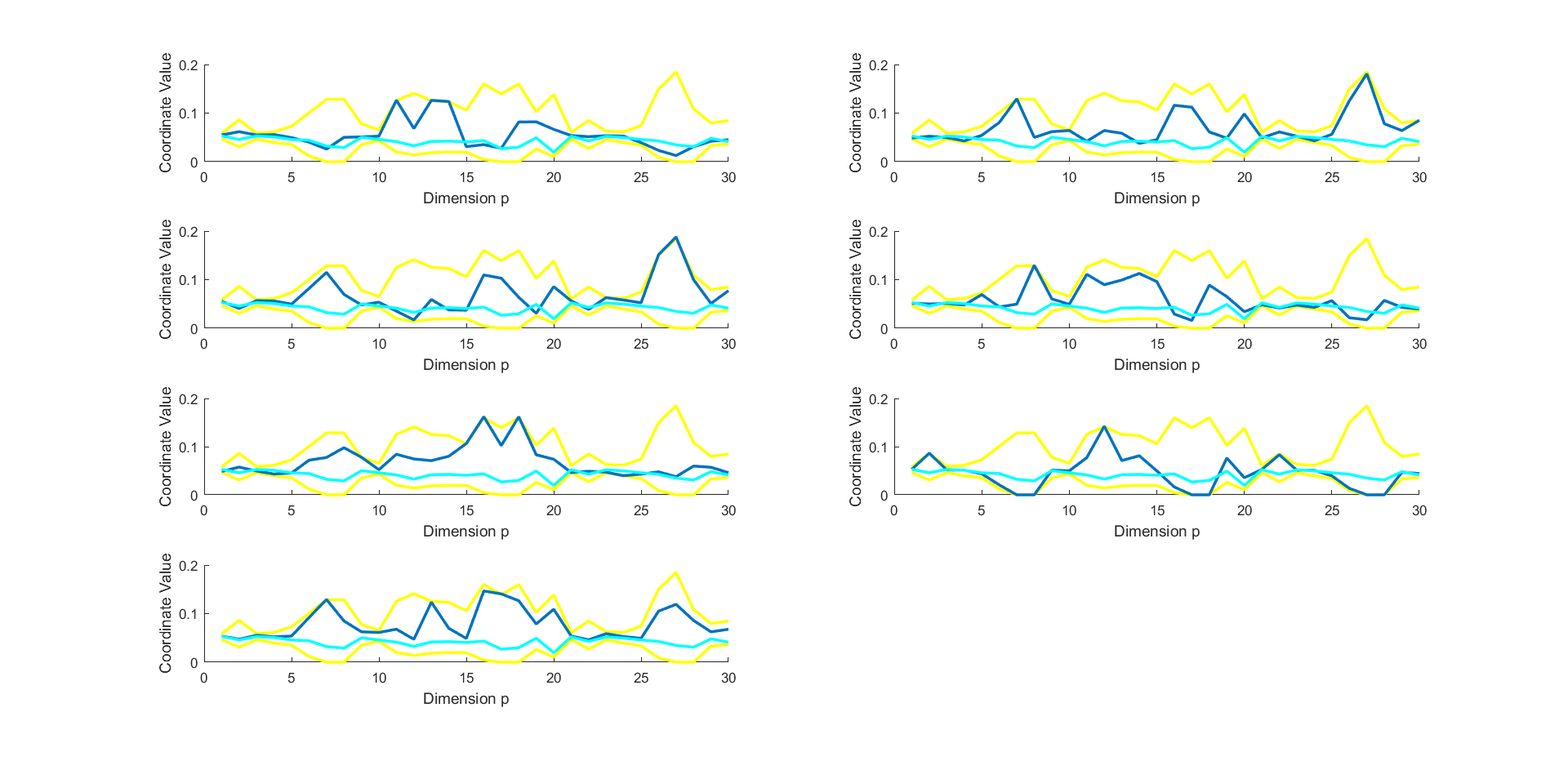}
     \caption{{\footnotesize RMD-Shrinkage detected outliers that belong to the $50\%$ of the most central data.}}            
     \label{outliersRMDv62}
\end{figure}

Table \ref{tab:computimereal} shows the computational times for each method in the task of detecting outliers with this example of real dataset. The results demonstrates that our competitors are much more slower than our proposal, except for \textit{COM} which has a similar computational time. 
\begin{table}[htbp]
  \centering
  \caption{Computational times for each methods with the WDBC dataset.}
    \begin{tabular}{l|llllll}
    Method & \multicolumn{1}{l}{MCD} & \multicolumn{1}{l}{Adj MCD} & \multicolumn{1}{l}{Kurtosis} & \multicolumn{1}{l}{OGK} & \multicolumn{1}{l}{COM} & \multicolumn{1}{l}{RMD-Shrinkage} \\
    \midrule
    Times in sec. & 12,155 & 12,3378 & 6,3077 & 3,5325 & 0,3534 & 0,3299 \\
    \end{tabular}%
  \label{tab:computimereal}%
\end{table}%

\section{Conclusions}

Correct detection of outliers in the multivariate case is well-known to be a very important task for thorough data analysis. In order to reach that goal properly, it is necessary to consider the shape of the data and its structure in the multivariate space. That is the reason why the Mahalanobis distance approach is frequently used for the task of identifying the outliers. Different robust Mahalanobis distances can be defined according to the selected robust location and dispersion estimators. There are various robust estimators in the literature that have been considered in this paper. A collection of different combinations of robust location and covariance matrix estimators based on the notion of shrinkage is proposed, in order to define with each combination a robust Mahalanobis distance for the outlier detection problem. The performance of the proposed \textit{RMD's} and the others from the literature is shown through a simulation study. It can be concluded that our competitors do not have a very desirable over-all behavior, especially in high dimension. The proposed \textit{RMD's} have the ability to discover outliers with high precision in the vast majority of cases in the simulations, with gaussian data and with skewed or heavy-tailed distributions. \textit{RMD-Shrinkage} is the most competitive version, as the simulation results showed. That is the reason why it is selected and some properties are investigated. The behavior under correlated and transformed data shows that \textit{RMD-Shrinkage} is approximately affine equivariant. With highly contaminated data it is shown that the approach has high breakdown value even in high dimension. There is also evidence of its inexpensive computational time. A real dataset example is also studied, in which the results bear out with the latter conclusions.
 
The results presented in this article emphasize the advantages
of using shrinkage estimators for the location and dispersion in the definition of a robust Mahalanobis distance. It remains to be examined whether the proposal could be improved by adapting the adjusted quantile to the proposed robust distances. It could also be an interesting matter to study, whether the use of the different definitions of ``depth'' in the literature (\cite{tukey1975mathematics}, \cite{liu1990notion}, \cite{serfling2002depth}, \cite{chen2009outlier}, \cite{agostinelli2011local}, \cite{paindaveine2013depth}), could improve the performance of the approach, as it is known that depth is a robust measure for location.

\section{Supplementary Material}

\subsection{Proofs}\label{prop1}

\subsubsection{Proof of Theorem 1.}

The optimization problem is:

\begin{equation}\label{eq:minsquarederrorMUCCM2}
\begin{matrix}
\textup{min}_{\nu_{\boldsymbol{\mu}},\eta} & E\left [ \left \| \boldsymbol{\hat{\mu}}_{Sh(CCM)} - \boldsymbol{\mu}  \right \|_2 ^2 \right ]\\ 
\textup{s.t.} & \boldsymbol{\hat{\mu}}_{Sh(CCM)} = (1-\eta) \boldsymbol{\hat{\mu}}_{CCM} + \eta \nu_{\boldsymbol{\mu}} \mathbf{e},
\end{matrix}
\end{equation}

\noindent
where $\left \| \mathbf{x} \right \|_2^2=\sum_{j=1}^p x_j^2$ and the associated inner product is: $\langle x,y \rangle =\sum_{j=1}^p x_j y_j$.\\

The objective function is equivalent to:

\begin{align*}
&E\left [ \left \| \boldsymbol{\hat{\mu}}_{Sh(CCM)} - \boldsymbol{\mu}  \right \|_2^2 \right ] \\
&= E\left [ \left \| (1-\eta) \boldsymbol{\hat{\mu}}_{CCM} + \eta \nu_{\boldsymbol{\mu}} \mathbf{e}  - \boldsymbol{\mu}  \right \|_2^2 \right ]\\  
 &=  (1-\eta)^2 E\left [ \left \| \boldsymbol{\hat{\mu}}_{CCM}-\boldsymbol{\mu}  \right \|_2^2 \right ] + \eta^2 \left \| \nu_{\boldsymbol{\mu}} \mathbf{e}  - \boldsymbol{\mu} \right \|_2^2 \\
&+ 2 E \left [ \left \langle   (1-\eta) (\boldsymbol{\hat{\mu}}_{CCM}-\boldsymbol{\mu}),  \eta (\nu_{\boldsymbol{\mu}} \mathbf{e}  - \boldsymbol{\mu}) \right \rangle \right ]
\end{align*}

The latter element in the above expression is equal to zero because $E(\boldsymbol{\hat{\mu}}_{CCM})=\boldsymbol{\mu}$ (see Chu [1995]). Then, the optimization problem (\ref{eq:minsquarederrorMUCCM2}) reduces to minimize:

\begin{align}\label{eq:expectedvaluesigmasMU}
E\left [ \left \| \boldsymbol{\hat{\mu}}_{Sh(CCM)} - \boldsymbol{\mu}  \right \|_2^2 \right ]&=(1-\eta)^2 E\left [ \left \| \boldsymbol{\hat{\mu}}_{CCM}-\boldsymbol{\mu}  \right \|_2^2 \right ]\\ \nonumber
&+ \eta^2 \left \| \nu_{\boldsymbol{\mu}} \mathbf{e}  - \boldsymbol{\mu} \right \|_2^2 
\end{align}

In order to find the optimal $\nu_{\boldsymbol{\mu}}$, it is necessary to minimize only the right element of the above expression.

\begin{equation*}
||\nu_{\boldsymbol{\mu}} \mathbf{e}  - \boldsymbol{\mu} ||_2^2 = \nu_{\boldsymbol{\mu}}^2 \left \| \mathbf{e}  \right \|_2^2 + \left \| \boldsymbol{\mu} \right \|_2^2 - 2 \nu_{\boldsymbol{\mu}} \left \langle \mathbf{e} ,\boldsymbol{\mu} \right \rangle
\end{equation*}

Then, with respect to the scaling parameter, the first order optimality condition give:

\begin{equation*}
0=2p\nu_{\boldsymbol{\mu}}  - 2  \left \langle \mathbf{e} ,\boldsymbol{\mu} \right \rangle= 2 \left( p \nu_{\boldsymbol{\mu}} - \sum_{j=1}^p \boldsymbol{\mu}_j \right)
\end{equation*}

Thus: 

\begin{equation*}
\nu_{\boldsymbol{\mu}}= \frac{1}{p}\sum_{j=1}^p \boldsymbol{\mu}_j 
\end{equation*}

Estimating $\boldsymbol{\mu}$ with $\boldsymbol{\hat{\mu}}_{CCM}$, we obtain:

$$\nu_{\boldsymbol{\mu}} = \frac{\hat{\boldsymbol{\mu}}_{CCM} \mathbf{e}}{p}$$

In (\ref{eq:expectedvaluesigmasMU}), with respect to the shrinkage intensity parameter $\eta$, the first order optimality condition give:

\begin{align*}
0&=2(1-\eta) E\left [ \left \| \boldsymbol{\hat{\mu}}_{CCM}-\boldsymbol{\mu} \right \|_2^2 \right ]  + 2\eta \left \|\nu_{\boldsymbol{\mu}} \mathbf{e}  - \boldsymbol{\mu}  \right \|_2^2\\
\end{align*}

Hence:

\begin{equation*}
\eta= \frac{E\left [ \left \| \hat{\boldsymbol{\mu}}_{CCM}-\boldsymbol{\mu} \right \|_2^2 \right ]}{E\left [ \left \| \hat{\boldsymbol{\mu}}_{CCM}- \nu_{\boldsymbol{\mu}}\mathbf{e}  \right \|_2^2 \right]  }
\end{equation*}

\subsubsection{Proof of Theorem 2.}

The optimization problem is:

\begin{equation}\label{eq:minsquarederrorMUMM2}
\begin{matrix}
\textup{min}_{\nu_{\boldsymbol{\mu}},\eta} & E\left [ \left \| \boldsymbol{\hat{\mu}}_{Sh(MM)} - \boldsymbol{\mu}  \right \|_2 ^2 \right ]\\ 
\textup{s.t.} & \boldsymbol{\hat{\mu}}_{Sh(MM)} = (1-\eta) \boldsymbol{\hat{\mu}}_{MM} + \eta \nu_{\boldsymbol{\mu}} \mathbf{e},
\end{matrix}
\end{equation}

\noindent
where $\left \| x \right \|_2^2=\sum_{j=1}^p x_j^2$.\\

Similarly to the previous demonstration, we can consider the following expression for the objective function:

\begin{align*}
E\left [ \left \| \boldsymbol{\hat{\mu}}_{Sh(MM)} - \boldsymbol{\mu}  \right \|_2^2 \right ]&= E\left [ \left \| (1-\eta) \boldsymbol{\hat{\mu}}_{MM} + \eta \nu_{\boldsymbol{\mu}} \mathbf{e}  - \boldsymbol{\mu}  \right \|_2^2 \right ]\\
&=  (1-\eta)^2 E\left [ \left \| \boldsymbol{\hat{\mu}}_{MM}-\boldsymbol{\mu}  \right \|_2^2 \right ] + \eta^2 \left \| \nu_{\boldsymbol{\mu}} \mathbf{e}  - \boldsymbol{\mu} \right \|_2^2 \\
&+ 2 E \left [ \left \langle   (1-\eta) (\boldsymbol{\hat{\mu}}_{MM}-\boldsymbol{\mu}),  \eta (\nu_{\boldsymbol{\mu}} \mathbf{e}  - \boldsymbol{\mu}) \right \rangle \right ]
\end{align*}

The expectation of the inner product is equal to zero because  Bose and Chaudhuri [1993] and Bose [1995] investigated the asymptotic distribution for the $L_1-$median, and they obtained the following result about the covariance matrix in presence of normality:

\begin{equation*}
\boldsymbol{\hat{\mu}}_{MM} \sim  N_p \left( \boldsymbol{\mu}, \frac{1}{n} \hat{A}^{-1} \hat{B} \hat{A}^{-1} \right),
\end{equation*}

Then, the optimization problem (\ref{eq:minsquarederrorMUMM2}) reduces to minimize:

\begin{align}\label{eq:expectedvaluesigmasMUMM}
E\left [ \left \| \boldsymbol{\hat{\mu}}_{Sh(MM)} - \boldsymbol{\mu}  \right \|_2^2 \right ]&=(1-\eta)^2 E\left [ \left \| \boldsymbol{\hat{\mu}}_{MM}-\boldsymbol{\mu}  \right \|_2^2 \right ] \\ \nonumber
&+ \eta^2 \left \| \nu_{\boldsymbol{\mu}} \mathbf{e}  - \boldsymbol{\mu} \right \|_2^2 
\end{align}

Then, the optimal parameter $\nu_{\boldsymbol{\mu}}$ can be found minimizing only the right element of the above expression, which is the only one depending on that parameter. 

\begin{equation*}
||\nu_{\boldsymbol{\mu}} \mathbf{e}  - \boldsymbol{\mu} ||_2^2 = \nu_{\boldsymbol{\mu}}^2 \left \| \mathbf{e}  \right \|_2^2 + \left \| \boldsymbol{\mu} \right \|_2^2 - 2 \nu_{\boldsymbol{\mu}} \left \langle \mathbf{e} ,\boldsymbol{\mu} \right \rangle
\end{equation*}

The associated first order optimality condition give:

\begin{equation*}
0=2p\nu_{\boldsymbol{\mu}}  - 2  \left \langle \mathbf{e} ,\boldsymbol{\mu} \right \rangle= 2 \left( p \nu_{\boldsymbol{\mu}} - \sum_{j=1}^p \boldsymbol{\mu}_j \right)
\end{equation*}

Therefore: 

\begin{equation*}
\nu_{\boldsymbol{\mu}}= \frac{1}{p}\sum_{j=1}^p \boldsymbol{\mu}_j 
\end{equation*}

In practice, we propose to estimate $\boldsymbol{\mu}$ with $\boldsymbol{\hat{\mu}}_{MM}$. Thus:

$$\nu_{\boldsymbol{\mu}} = \frac{\hat{\boldsymbol{\mu}}_{MM} \mathbf{e}}{p}$$

With respect to the shrinkage intensity parameter $\eta$, the first order optimality condition associated to (\ref{eq:expectedvaluesigmasMUMM}), give:

\begin{align*}
0&=2(1-\eta) E\left [ \left \| \boldsymbol{\hat{\mu}}_{MM}-\boldsymbol{\mu} \right \|_2^2 \right ]  + 2\eta \left \|\nu_{\boldsymbol{\mu}} \mathbf{e}  - \boldsymbol{\mu}  \right \|_2^2\\
\end{align*}

Hence:

\begin{equation*}
\eta= \frac{E\left [ \left \| \hat{\boldsymbol{\mu}}_{MM}-\boldsymbol{\mu} \right \|_2^2 \right ]}{E\left [ \left \| \hat{\boldsymbol{\mu}}_{MM}- \nu_{\boldsymbol{\mu}}\mathbf{e}  \right \|_2^2 \right]  }
\end{equation*}

\subsubsection{Proof of Theorem 3.}

The optimization problem is:

\begin{equation}\label{eq:minsquarederrorSIGMA2}
\begin{matrix}
\textup{min}_{\nu_{\Sigma},\eta} & E\left [ \left \| \hat{\Sigma}_{Sh} - \Sigma  \right \| ^2 \right ]\\ 
\textup{s.t.} & \hat{\Sigma}_{Sh} = (1-\eta) \hat{S}_{CCM} + \eta \nu_{\Sigma} I,
\end{matrix}
\end{equation}

\noindent
where $\left \|A\right \|^2=trace(AA^T)/p$, and the associated inner product is $\langle A_1, A_2 \rangle=trace(A_1 A_2^T)/p$.

Analogous to the previous Propositions, the objective function in the above minimization problem (\ref{eq:minsquarederrorSIGMA2}) can be seen as:

\begin{align*}
&E\left [ \left \| \hat{\Sigma}_{Sh} - \Sigma  \right \| ^2 \right ] \\
&= E\left [ \left \| (1-\eta) \hat{S}_{CCM} + \eta \nu_{\Sigma} I - \Sigma  \right \| ^2 \right ]\\  
 &=  (1-\eta)^2 E\left [ \left \| \hat{S}_{CCM}-\Sigma  \right \| ^2 \right ] + \eta^2 \left \| \nu_{\Sigma} I - \Sigma \right \|^2 \\
&+ 2 E \left [ \left \langle   (1-\eta) (\hat{S}_{CCM}-\Sigma),  \eta (\nu_{\Sigma} I - \Sigma) \right \rangle \right ]
\end{align*}

In this case, note that the latter element in the above expression is equal to zero because $E(\hat{S}_{CCM})=\Sigma$. Hence, the optimization problem (\ref{eq:minsquarederrorSIGMA2}) reduces to minimize the following  expression:

\begin{equation}\label{eq:expectedvaluesigmas}
E\left [ \left \| \hat{\Sigma}_{Sh} - \Sigma  \right \| ^2 \right ]=(1-\eta)^2 E\left [ \left \| \hat{S}_{CCM}-\Sigma  \right \| ^2 \right ] + \eta^2 \left \| \nu_{\Sigma} I - \Sigma \right \|^2 
\end{equation}

The optimal $\nu_{\Sigma}$ can be obtained by minimizing only the right element of the above expression, because it is the only one depending on that parameter. Also, note that:

\begin{equation}
||\nu_{\Sigma} I - \Sigma ||^2 = \nu_{\Sigma}^2 \left \| I \right \|^2 + \left \| \Sigma \right \|^2 - 2 \nu_{\Sigma} \left \langle I,\Sigma \right \rangle
\end{equation}

Then, the first order optimality condition with respect to the scaling parameter, give: 

\begin{align*}
0&=2\nu_{\Sigma}  - 2  \left \langle I,\Sigma \right \rangle\\
\nu_{\Sigma}&=\left \langle I,\Sigma \right \rangle=trace(\Sigma I^T)/p 
\end{align*}

Therefore:
$$\nu_{\Sigma}= trace(\Sigma)/p$$

In practice, we propose to estimate $\Sigma$ with $\hat{S}_{CCM}$, thus:

$$\nu_{\Sigma}=trace(\hat{S}_{CCM})/p$$

In (\ref{eq:expectedvaluesigmas}), with respect to the shrinkage intensity parameter $\eta$, the first order optimality condition give:

\begin{equation*}
\eta= \frac{E\left [ \left \| \hat{S}_{CCM}-\Sigma \right \|^2 \right ]}{E\left [ \left \| \hat{S}_{CCM}- \nu_{\Sigma}I \right \|^2 \right ]  }
\end{equation*}

\newpage

\subsection{Tables}

\subsubsection{Normal distribution}

\small

Table \ref{tab:0contam} shows the false detection rates when there is no contamination. The Tables \ref{tab:c_normallambda5}-\ref{tab:c_normallambda10} show the correct detection rates (c) and Tables \ref{tab:f_normallambda5}-\ref{tab:f_normallambda10} the false detection rates (f), for each method, corresponding to the simulations with multivariate Normal distribution, for contamination levels $\alpha=0.1,0.2,0.3$, dimension $p=5,10,30$, distance of the outliers $\delta=5,10$ and concentration of the contamination $\lambda=0.1,1$.

\begin{table}[htbp]
  \centering
  \caption{False detection rates with Normal distribution $\alpha=0$.\vspace{3cm}}
  \resizebox{15cm}{!} {
    \begin{tabular}{c|ccccccccccc}
	    &     &  &   &     &     &   &   &   &   &   &   \\   
   $ p $   & MCD   & Adj MCD & Kurtosis & OGK   & COM   & RMDv1 & RMDv2 & RMDv3 & RMDv4 & RMDv5 & RMDv6 \\
    \midrule
    5     & 0,06440 & 0,04640 & 0,02630 & 0,08120 & 0,00480 & 0,03140 & 0,02820 & 0,02770 & 0,03100 & 0,02900 & 0,00316 \\
    10    & 0,11760 & 0,09830 & 0,07110 & 0,09580 & 0,00217 & 0,00245 & 0,00222 & 0,00215 & 0,00172 & 0,00161 & 0,00160 \\
    30    & 0,06276 & 0,03922 & 0,00804 & 0,09084 & 0,00008 & 0,00003 & 0,00003 & 0,00003 & 0,00003 & 0,00002 & 0,00001 \\
    \end{tabular}%
    }
  \label{tab:0contam}%
\end{table}%

\begin{table}[htbp]
  \centering
  \caption{Correct detection rates with Normal distribution.}
   \resizebox{15cm}{!} {
    \begin{tabular}{c|cccccccccccc}
       &   &       &       &       &       &       &       &       &       &       &       &  \\
    $\delta=5$ &   &  $\lambda=0.1$     &       &       &       &       &       &       &       &       &       &  \\
    \midrule
    $p$     & $\alpha$ & MCD   & Adj MCD & Kurtosis & OGK   & COM   & RMDv1 & RMDv2 & RMDv3 & RMDv4 & RMDv5 & RMDv6 \\
    \midrule
    5     & 0.1   & 1     & 1     & 0,9000 & 1     & 1     & 1     & 1     & 1     & 1     & 1     & 1 \\
          & 0.2   & 0,8700 & 0,8700 & 0,5100 & 0,9500 & 0,9941 & 1     & 1     & 1     & 1     & 1     & 1 \\
          & 0.3   & 0,0600 & 0,0600 & 0,9800 & 0,1500 & 0,5719 & 0,8766 & 0,8782 & 0,8782 & 0,9146 & 0,9090 & 0,9130 \\
    \midrule
    10    & 0.1   & 0,9900 & 0,9900 & 0,8600 & 1     & 1     & 1     & 1     & 1     & 1     & 1     & 1 \\
          & 0.2   & 0,2800 & 0,2800 & 0,4600 & 0,9416 & 1     & 1     & 1     & 1     & 1     & 1     & 1 \\
          & 0.3   & 0     & 0     & 0,9900 & 0,1612 & 0,7205 & 0,8774 & 0,8747 & 0,8750 & 0,9711 & 0,9672 & 0,9711 \\
    \midrule
    30    & 0.1   & 0,1900 & 0,1900 & 1     & 1     & 1     & 1     & 1     & 1     & 1     & 1     & 1 \\
          & 0.2   & 0     & 0     & 0,1   & 1     & 1     & 1     & 1     & 1     & 1     & 1     & 1 \\
          & 0.3   & 0     & 0     & 0,6100 & 0,0100 & 0,9407 & 0,5308 & 0,5275 & 0,5286 & 0,9990 & 0,9988 & 0,9991 \\
          \midrule
    $\delta=5$ & $\alpha$ &  $\lambda=1$     &       &       &       &       &       &       &       &       &       &  \\
    \midrule
    5     & 0.1   & 1     & 1     & 1     & 1     & 1     & 1     & 1     & 1     & 1     & 1     & 1 \\
          & 0.2   & 0,8578 & 0,8486 & 0,9602 & 0,9654 & 0,9975 & 1     & 1     & 1     & 1     & 1     & 1 \\
          & 0.3   & 0,1955 & 0,1664 & 0,9336 & 0,5792 & 0,8735 & 0,8935 & 0,8947 & 0,8938 & 0,8740 & 0,8698 & 0,8755 \\
    \midrule
    10    & 0.1   & 1     & 1     & 1     & 1     & 1     & 1     & 1     & 1     & 1     & 1     & 1 \\
          & 0.2   & 0,9016 & 0,8935 & 0,7212 & 0,9993 & 1     & 1     & 1     & 1     & 1     & 1     & 1 \\
          & 0.3   & 0,2375 & 0,2080 & 0,5935 & 0,6108 & 0,9505 & 0,8846 & 0,8838 & 0,8838 & 0,9608 & 0,9581 & 0,9621 \\
    \midrule
    30    & 0.1   & 1     & 1     & 0,8816 & 1     & 1     & 1     & 1     & 1     & 1     & 1     & 1 \\
          & 0.2   & 0,4461 & 0,4232 & 0,0154 & 1     & 1     & 1     & 1     & 1     & 1     & 1     & 1 \\
          & 0.3   & 0,0823 & 0,0532 & 0,1483 & 0,9772 & 0,9990 & 0,7142 & 0,7035 & 0,7059 & 1     & 1     & 1 \\
    \end{tabular}%
    }
    \label{tab:c_normallambda5}%
\end{table}%

\begin{table}[htbp]
  \centering
  \caption{Correct detection rates with Normal distribution.}
   \resizebox{15cm}{!} {
    \begin{tabular}{c|cccccccccccc}
          &   &       &       &       &       &       &       &       &       &       &       &  \\
    $\delta=10$ &  &   $\lambda=0.1$    &       &       &       &       &       &       &       &       &       &  \\
    \midrule
    $p$     & $\alpha$ & MCD   & Adj MCD & Kurtosis & OGK   & COM   & RMDv1 & RMDv2 & RMDv3 & RMDv4 & RMDv5 & RMDv6 \\
    \midrule
    5     & 0.1   & 1     & 1     & 0,9200 & 1     & 1     & 1     & 1     & 1     & 1     & 1     & 1 \\
          & 0.2   & 0,8200 & 0,8200 & 0,6500 & 1     & 1     & 1     & 1     & 1     & 1     & 1     & 1 \\
          & 0.3   & 0,1000 & 0,1000 & 1     & 0,7400 & 1     & 1     & 1     & 1     & 1     & 1     & 1 \\
    \midrule
    10    & 0.1   & 1     & 1     & 0,9200 & 1     & 1     & 1     & 1     & 1     & 1     & 1     & 1 \\
          & 0.2   & 0,7200 & 0,7200 & 0,4400 & 1     & 1     & 1     & 1     & 1     & 1     & 1     & 1 \\
          & 0.3   & 0,0500 & 0,0500 & 0,9700 & 0,7500 & 1     & 1     & 1     & 1     & 1     & 1     & 1 \\
    \midrule
    30    & 0.1   & 0,8800 & 0,8800 & 1     & 1     & 1     & 1     & 1     & 1     & 1     & 1     & 1 \\
          & 0.2   & 0     & 0     & 0,1200 & 1     & 1     & 1     & 1     & 1     & 1     & 1     & 1 \\
          & 0.3   & 0     & 0     & 0,5400 & 0,9300 & 1     & 1     & 1     & 1     & 1     & 1     & 1 \\
    \midrule
    $\delta=10$ &  $\alpha$ &    $\lambda=1$   &       &       &       &       &       &       &       &       &       &  \\
    \midrule
    5     & 0.1   & 1     & 1     & 1     & 1     & 1     & 1     & 1     & 1     & 1     & 1     & 1 \\
          & 0.2   & 0,8480 & 0,8465 & 0,9900 & 1     & 1     & 1     & 1     & 1     & 1     & 1     & 1 \\
          & 0.3   & 0,2190 & 0,1976 & 0,9307 & 0,9591 & 0,9991 & 1     & 1     & 1     & 1     & 1     & 1 \\
    \midrule
    10    & 0.1   & 1     & 1     & 0,9800 & 1     & 1     & 1     & 1     & 1     & 1     & 1     & 1 \\
          & 0.2   & 0,8623 & 0,8548 & 0,6558 & 1     & 1     & 1     & 1     & 1     & 1     & 1     & 1 \\
          & 0.3   & 0,2280 & 0,2046 & 0,4618 & 0,9911 & 1     & 1     & 1     & 1     & 1     & 1     & 1 \\
    \midrule
    30    & 0.1   & 1     & 1     & 0,8919 & 1     & 1     & 1     & 1     & 1     & 1     & 1     & 1 \\
          & 0.2   & 0,4879 & 0,4654 & 0,0125 & 1     & 1     & 1     & 1     & 1     & 1     & 1     & 1 \\
          & 0.3   & 0,0810 & 0,0509 & 0,1087 & 1     & 1     & 1     & 1     & 1     & 1     & 1     & 1 \\
    \end{tabular}%
    }
  \label{tab:c_normallambda10}%
\end{table}%

\begin{table}[htbp]
  \centering
  \caption{False detection rates with Normal distribution $\delta=5$.}
  \resizebox{15cm}{!} {
    \begin{tabular}{c|cccccccccccc}
          &   &       &       &       &       &       &       &       &       &       &       &  \\
     &  &   $\lambda=0.1$     &       &       &       &       &       &       &       &       &       &  \\
    \midrule
    $p$     & $\alpha$ & MCD   & Adj MCD & Kurtosis & OGK   & COM   & RMDv1 & RMDv2 & RMDv3 & RMDv4 & RMDv5 & RMDv6 \\
    \midrule
    5     & 0.1   & 0,0327 & 0,0184 & 0,0336 & 0,0640 & 0,0040 & 0,0080 & 0,0070 & 0,0073 & 0,0069 & 0,0067 & 0,0076 \\
          & 0.2   & 0,0265 & 0,0171 & 0,0512 & 0,0600 & 0,0028 & 0,0015 & 0,0015 & 0,0012 & 0,0013 & 0,0013 & 0,0013 \\
          & 0.3   & 0,1504 & 0,1247 & 0,0373 & 0,1641 & 0,0015 & 0     & 0     & 0     & 0     & 0     & 0 \\
    \midrule
    10    & 0.1   & 0,0735 & 0,0529 & 0,1167 & 0,0823 & 0,0027 & 0,0055 & 0,0057 & 0,0048 & 0,0025 & 0,0024 & 0,0027 \\
          & 0.2   & 0,1566 & 0,1330 & 0,2803 & 0,0667 & 0,0023 & 0,0002 & 0,0002 & 0,0001 & 0,0001 & 0,0001 & 0,0001 \\
          & 0.3   & 0,2485 & 0,2206 & 0,0767 & 0,2502 & 0,0010 & 0     & 0     & 0     & 0     & 0     & 0 \\
    \midrule
    30    & 0.1   & 0,0767 & 0,0507 & 0,0078 & 0,0699 & 5E-05 & 0,0007 & 0,0006 & 0,0006 & 0,0001 & 4,4E-05 & 0,0001 \\
          & 0.2   & 0,1079 & 0,0784 & 0,0565 & 0,0552 & 3E-05 & 0     & 0     & 0     & 0     & 0     & 0 \\
          & 0.3   & 0,1491 & 0,1150 & 0,0547 & 0,5030 & 3E-05 & 0     & 0     & 0     & 0     & 0     & 0 \\
    \midrule
    & $\alpha$ &  $\lambda=1$      &       &       &       &       &       &       &       &       &       &  \\
    \midrule
    5     & 0.1   & 0,0339 & 0,0198 & 0,0341 & 0,0636 & 0,0033 & 0,0084 & 0,0073 & 0,0069 & 0,0075 & 0,0071 & 0,0077 \\
          & 0.2   & 0,0110 & 0,0055 & 0,0347 & 0,0507 & 0,0017 & 0,0014 & 0,0012 & 0,0012 & 0,0010 & 0,0009 & 0,0012 \\
          & 0.3   & 0,0374 & 0,0260 & 0,0379 & 0,0383 & 0,0001 & 0     & 0     & 0     & 0     & 0     & 0 \\
    \midrule
    10    & 0.1   & 0,0704 & 0,0505 & 0,0917 & 0,0842 & 0,0023 & 0,0047 & 0,0042 & 0,0041 & 0,0027 & 0,0022 & 0,0029 \\
          & 0.2   & 0,0270 & 0,0171 & 0,0860 & 0,0610 & 0,0010 & 0,0002 & 0,0002 & 0,0002 & 0,0001 & 0,0001 & 0,0001 \\
          & 0.3   & 0,0851 & 0,0706 & 0,0782 & 0,0457 & 0,0003 & 0     & 0     & 0     & 0     & 0     & 0 \\
    \midrule
    30    & 0.1   & 0,0347 & 0,0115 & 0,0081 & 0,0713 & 0,0001 & 0,0008 & 0,0007 & 0,0007 & 0,0001 & 0,0001 & 0,0001 \\
          & 0.2   & 0,0357 & 0,0198 & 0,0066 & 0,0544 & 0     & 0     & 0     & 0     & 0     & 0     & 0 \\
          & 0.3   & 0,0552 & 0,0343 & 0,0077 & 0,0366 & 0     & 0     & 0     & 0     & 0     & 0     & 0 \\
    \end{tabular}%
    }
  \label{tab:f_normallambda5}%
\end{table}%

\begin{table}[htbp]
  \centering
  \caption{False detection rates with Normal distribution $\delta=10$.}
   \resizebox{15cm}{!} {
    \begin{tabular}{c|cccccccccccc}
          &   &       &       &       &       &       &       &       &       &       &       &  \\
      &   &   $\lambda=0.1$    &       &       &       &       &       &       &       &       &       &  \\
    \midrule
   $ p$    & $\alpha$ & MCD   & Adj MCD & Kurtosis & OGK   & COM   & RMDv1 & RMDv2 & RMDv3 & RMDv4 & RMDv5 & RMDv6 \\
    \midrule
    5     & 0.1   & 0,0309 & 0,0154 & 0,0301 & 0,0681 & 0,0022 & 0,0090 & 0,0076 & 0,0066 & 0,0066 & 0,0058 & 0,0069 \\
          & 0.2   & 0,0337 & 0,0248 & 0,0414 & 0,0485 & 0,0036 & 0,0005 & 0,0004 & 0,0004 & 0,0011 & 0,0008 & 0,0007 \\
          & 0.3   & 0,1434 & 0,1183 & 0,0309 & 0,0603 & 0,0031 & 0     & 0     & 0     & 0     & 0     & 0 \\
    \midrule
    10    & 0.1   & 0,0665 & 0,0464 & 0,1147 & 0,0772 & 0,0017 & 0,0041 & 0,0047 & 0,0042 & 0,0034 & 0,0030 & 0,0035 \\
          & 0.2   & 0,0809 & 0,0647 & 0,2827 & 0,0606 & 0,0022 & 0,0002 & 0,0001 & 0,0001 & 0     & 0     & 0 \\
          & 0.3   & 0,2353 & 0,2088 & 0,0934 & 0,0895 & 0,0008 & 0     & 0     & 0     & 0     & 0     & 0 \\
    \midrule
    30    & 0.1   & 0,0415 & 0,0174 & 0,0081 & 0,0715 & 4E-05 & 0,0011 & 0,0010 & 0,0011 & 0,0001 & 0,0001 & 0,0001 \\
          & 0.2   & 0,1072 & 0,0776 & 0,0481 & 0,0561 & 3E-05 & 0     & 0     & 0     & 0     & 0     & 0 \\
          & 0.3   & 0,1500 & 0,1163 & 0,0578 & 0,0623 & 0,0001 & 0     & 0     & 0     & 0     & 0     & 0 \\
    \midrule
    & $\alpha$  &  $\lambda=1$     &       &       &       &       &       &       &       &       &       &  \\
    \midrule
    5     & 0.1   & 0,0332 & 0,0174 & 0,0297 & 0,0682 & 0,0021 & 0,0076 & 0,0068 & 0,0069 & 0,0058 & 0,0052 & 0,0062 \\
          & 0.2   & 0,0136 & 0,0058 & 0,0352 & 0,0535 & 0,0023 & 0,0011 & 0,0012 & 0,0011 & 0,0008 & 0,0008 & 0,0007 \\
          & 0.3   & 0,0385 & 0,0289 & 0,0372 & 0,0397 & 0,0007 & 0     & 0     & 0     & 0,0001 & 0     & 0 \\
    \midrule
    10    & 0.1   & 0,0668 & 0,0469 & 0,0907 & 0,0771 & 0,0020 & 0,0036 & 0,0036 & 0,0031 & 0,0025 & 0,0021 & 0,0026 \\
          & 0.2   & 0,0278 & 0,0170 & 0,0930 & 0,0629 & 0,0009 & 0,0002 & 0,0001 & 0,0002 & 0     & 0     & 0 \\
          & 0.3   & 0,0856 & 0,0694 & 0,0685 & 0,0440 & 0,0004 & 0     & 0     & 0     & 0     & 0     & 0 \\
    \midrule
    30    & 0.1   & 0,0351 & 0,0122 & 0,0077 & 0,0735 & 4E-05 & 0,0009 & 0,0008 & 0,0009 & 0,0001 & 2,21E-05 & 0,0001 \\
          & 0.2   & 0,0334 & 0,0181 & 0,0053 & 0,0535 & 0     & 0     & 0     & 0     & 0     & 0     & 0 \\
          & 0.3   & 0,0577 & 0,0368 & 0,0085 & 0,0388 & 0     & 0     & 0     & 0     & 0     & 0     & 0 \\
    \end{tabular}%
    }
  \label{tab:f_normallambda10}%
\end{table}%

\newpage

\subsubsection{Computational times}

\begin{table}[htbp]
  \centering
\caption{Computational times with Normal data $\delta=5$ and $\lambda=1$.}
\resizebox{13cm}{!} {
    \begin{tabular}{c|ccccccc}
        &    &     &   & &     &     &   \\
    $ p  $   & $\alpha$ & MCD   & Adj MCD & Kurtosis & OGK   & COM   & RMD-Shrinkage \\
    \midrule
    5     & 0.1   & 0,8547 & 0,8078 & 0,0959 & 0,0181 & 0,0070 & 0,0029 \\
          & 0.2   & 1,2146 & 0,7129 & 0,0763 & 0,0186 & 0,0061 & 0,0034 \\
          & 0.3   & 1,0064 & 0,7544 & 0,0612 & 0,0176 & 0,0063 & 0,0025 \\
          & \textbf{Mean} & \textbf{1,0252} & \textbf{0,7584} & \textbf{0,0778} & \textbf{0,0181} & \textbf{0,0065} & \textbf{0,0030} \\
    \midrule
    10    & 0.1   & 1,0090 & 1,1250 & 0,1592 & 0,0793 & 0,0113 & 0,0047 \\
          & 0.2   & 1,0135 & 1,0448 & 0,1679 & 0,0623 & 0,0100 & 0,0025 \\
          & 0.3   & 1,0335 & 1,0595 & 0,1515 & 0,0612 & 0,0091 & 0,0009 \\
          & \textbf{Mean} & \textbf{1,0187} & \textbf{1,0765} & \textbf{0,1595} & \textbf{0,0676} & \textbf{0,0101} & \textbf{0,0027} \\
    \midrule
    30    & 0.1   & 6,5263 & 6,2629 & 0,4788 & 0,9623 & 0,2530 & 0,2752 \\
          & 0.2   & 6,1268 & 6,2031 & 0,5737 & 0,8317 & 0,1700 & 0,2139 \\
          & 0.3   & 5,9068 & 6,1767 & 0,5034 & 0,9472 & 0,1791 & 0,2101 \\
          & \textbf{Mean} & \textbf{6,1866} & \textbf{6,2142} & \textbf{0,5186} & \textbf{0,9137} & \textbf{0,2007} & \textbf{0,2331} \\
    \end{tabular}%
    }
  \label{tab:tiempos2}%
\end{table}%

\begin{table}[htbp]
\caption{Computational times with Normal data $\delta=10$ and $\lambda=0.1$.}
\resizebox{13cm}{!} {
    \begin{tabular}{c|ccccccc}
            &    &     &   & &     &     &   \\
$ p  $   & $\alpha$ & MCD   & Adj MCD & Kurtosis & OGK   & COM   & RMD-Shrinkage \\
    \midrule
    \multicolumn{1}{c|}{5} & 0.1   & 0,7704 & 0,7425 & 0,1090 & 0,0195 & 0,0107 & 0,0026 \\
    \multicolumn{1}{c|}{} & 0.2   & 0,6878 & 0,6534 & 0,0573 & 0,0195 & 0,0079 & 0,0033 \\
    \multicolumn{1}{c|}{} & 0.3   & 0,6990 & 0,7434 & 0,0294 & 0,0268 & 0,0072 & 0,0054 \\
    \multicolumn{1}{c|}{} & \textbf{Mean} & \textbf{0,7191} & \textbf{0,7131} & \textbf{0,0652} & \textbf{0,0219} & \textbf{0,0086} & \textbf{0,0038} \\
    \midrule
    \multicolumn{1}{c|}{10} & 0.1   & 1,0631 & 1,1115 & 0,1297 & 0,0824 & 0,0101 & 0,0066 \\
    \multicolumn{1}{c|}{} & 0.2   & 1,1940 & 0,9625 & 0,1179 & 0,0719 & 0,0080 & 0,0041 \\
    \multicolumn{1}{c|}{} & 0.3   & 1,0673 & 0,9902 & 0,0756 & 0,0663 & 0,0097 & 0,0047 \\
    \multicolumn{1}{c|}{} & \textbf{Mean} & \textbf{1,1081} & \textbf{1,0214} & \textbf{0,1077} & \textbf{0,0735} & \textbf{0,0093} & \textbf{0,0051} \\
    \midrule
    \multicolumn{1}{c|}{30} & 0.1   & 6,0350 & 6,0913 & 0,7347 & 0,8205 & 0,1701 & 0,1697 \\
    \multicolumn{1}{c|}{} & 0.2   & 6,4506 & 6,1627 & 0,7385 & 0,7491 & 0,1837 & 0,1572 \\
    \multicolumn{1}{c|}{} & 0.3   & 6,2114 & 6,1146 & 1,1310 & 0,7342 & 0,1714 & 0,1281 \\
          & \textbf{Mean} & \textbf{6,2324} & \textbf{6,1229} & \textbf{0,8681} & \textbf{0,7679} & \textbf{0,1750} & \textbf{0,1517} \\
    \end{tabular}%
    }
  \label{tab:tiempos3}%
\end{table}%

\begin{table}[htbp]
  \caption{Computational times with Normal data $\delta=10$ and $\lambda=1$.}
  \resizebox{13cm}{!} {
    \begin{tabular}{c|ccccccc}
            &    &     &   & &     &     &   \\
$ p  $   & $\alpha$ & MCD   & Adj MCD & Kurtosis & OGK   & COM   & RMD-Shrinkage \\
    \midrule
    5     & 0.1   & 0,7956 & 0,8301 & 0,0814 & 0,0176 & 0,0078 & 0,0037 \\
          & 0.2   & 0,7939 & 0,7684 & 0,0836 & 0,0189 & 0,0124 & 0,0019 \\
          & 0.3   & 0,8614 & 0,7207 & 0,0662 & 0,0183 & 0,0049 & 0,0014 \\
          & \textbf{Mean} & \textbf{0,8170} & \textbf{0,7731} & \textbf{0,0770} & \textbf{0,0183} & \textbf{0,0084} & \textbf{0,0023} \\
    \midrule
    10    & 0.1   & 0,9990 & 1,0609 & 0,1350 & 0,0634 & 0,0117 & 0,0047 \\
          & 0.2   & 1,0917 & 1,1028 & 0,1613 & 0,0682 & 0,0093 & 0,0049 \\
          & 0.3   & 1,0111 & 1,1860 & 0,1610 & 0,0766 & 0,0089 & 0,0025 \\
          & \textbf{Mean} & \textbf{1,0340} & \textbf{1,1166} & \textbf{0,1524} & \textbf{0,0694} & \textbf{0,0100} & \textbf{0,0040} \\
    \midrule
    30    & 0.1   & 5,7161 & 5,6622 & 0,5731 & 0,7563 & 0,1693 & 0,1220 \\
          & 0.2   & 5,6394 & 5,6993 & 0,5821 & 0,7191 & 0,1654 & 0,1083 \\
          & 0.3   & 5,7104 & 5,7975 & 0,9465 & 0,7193 & 0,1561 & 0,1138 \\
          & \textbf{Mean} & \textbf{5,6886} & \textbf{5,7197} & \textbf{0,7006} & \textbf{0,7316} & \textbf{0,1636} & \textbf{0,1147} \\
    \end{tabular}%
    }
  \label{tab:tiempos4}%
\end{table}%

\newpage

\subsection{Multivariate Student-t distribution with 3 degrees of freedom}

Table \ref{tab:0contamt} shows the false detection rates when there is no contamination. Tables \ref{tab:c_studentlambda5}-\ref{tab:c_studentlambda10} show the correct detection rates (c) and Tables \ref{tab:f_studentlambda5}-\ref{tab:f_studentlambda10} the false detection rates (f), for each method, corresponding to the simulations with multivariate Student-t distribution with 3 degrees of freedom, for contamination levels $\alpha=0.1,0.2,0.3$, dimension $p=5,10,30$, distance of the outliers $\delta=5,10$ and concentration of the contamination $\lambda=0.1,1$. 

\begin{table}[htbp]
  \centering
  \caption{False detection rates with Student-t distribution with 3 d.f, $\alpha=0$.}
  \resizebox{15cm}{!} {
    \begin{tabular}{c|ccccccccccc}
       &     &   &   &     &     &   &   &   &   &   &   \\
   $ p$     & MCD   & Adj MCD & Kurtosis & OGK   & COM   & RMDv1 & RMDv2 & RMDv3 & RMDv4 & RMDv5 & RMDv6 \\
    \midrule
    5     & 0,12860 & 0,11060 & 0,17670 & 0,21320 & 0,07730 & 0,20290 & 0,20070 & 0,20030 & 0,17950 & 0,17660 & 0,18370 \\
    10    & 0,17560 & 0,15640 & 0,31610 & 0,24330 & 0,08410 & 0,19040 & 0,19550 & 0,19570 & 0,11780 & 0,11220 & 0,12760 \\
    30    & 0,15820 & 0,13530 & 0,20380 & 0,28760 & 0,08270 & 0,16660 & 0,16430 & 0,16490 & 0,11220 & 0,11130 & 0,11400 \\
    \end{tabular}%
    }
  \label{tab:0contamt}%
\end{table}%

\begin{table}[htbp]
  \centering
  \caption{Correct detection rates with Student-t with 3 d.f. and $\delta=5$.}
   \resizebox{15cm}{!} {
    \begin{tabular}{c|cccccccccccc}
           &     &   &   &     &     &   &   &   &   &   &   \\
          & $\lambda= 0.1 $&       &       &       &       &       &       &       &       &       &       &  \\
    \midrule
   $ p $    & $\alpha$ & MCD   & Adj MCD & Kurtosis & OGK   & COM   & RMDv1 & RMDv2 & RMDv3 & RMDv4 & RMDv5 & RMDv6 \\
    \midrule
    5     & 0.1   & 0,9300 & 0,9300 & 0,4900 & 1     & 1     & 1     & 1     & 1     & 1     & 1     & 1 \\
          & 0.2   & 0,1800 & 0,1800 & 0,2902 & 0,5300 & 0,6872 & 1     & 1     & 1     & 0,9927 & 0,9915 & 0,9915 \\
          & 0.3   & 0,0007 & 0,0007 & 0,6890 & 0,0007 & 0,0411 & 0,6573 & 0,6446 & 0,6441 & 0,6878 & 0,6758 & 0,7072 \\
    \midrule
    10    & 0.1   & 0,5107 & 0,5107 & 0,3407 & 0,9800 & 1     & 1     & 1     & 1     & 1     & 1     & 1 \\
          & 0.2   & 0,0009 & 0,0009 & 0,1809 & 0,3709 & 0,6576 & 1,0000 & 1,0000 & 1,0000 & 0,9997 & 0,9997 & 1 \\
          & 0.3   & 0,0100 & 0,0100 & 0,6200 & 0,0203 & 0,0400 & 0,6500 & 0,6525 & 0,6522 & 0,8021 & 0,7947 & 0,8137 \\
    \midrule
    30    & 0.1   & 0,0003 & 0,0003 & 0,2000 & 1     & 1     & 1     & 1     & 1     & 1     & 1     & 1 \\
          & 0.2   & 0,0004 & 0,0004 & 0,1703 & 0,1905 & 0,6055 & 1     & 1     & 1     & 1     & 1     & 1 \\
          & 0.3   & 0     & 0     & 1     & 0,0002 & 0     & 0,1689 & 0,1638 & 0,1649 & 0,7437 & 0,7417 & 0,7628 \\
    \midrule
   $ p $    & $\alpha$ & $\lambda=1$ &       &       &       &       &       &       &       &       &       &  \\
    \midrule
    5     & 0.1   & 1     & 1     & 1     & 1     & 1     & 1     & 1     & 1     & 1     & 1     & 1 \\
          & 0.2   & 0,7445 & 0,7263 & 0,8974 & 0,9177 & 0,9155 & 0,9970 & 0,9952 & 0,9952 & 0,9922 & 0,9919 & 0,9924 \\
          & 0.3   & 0,1844 & 0,1591 & 0,8421 & 0,4498 & 0,4732 & 0,8442 & 0,8485 & 0,8457 & 0,8411 & 0,8360 & 0,8377 \\
    \midrule
    10    & 0.1   & 0,9826 & 0,9819 & 0,9492 & 1     & 1     & 1     & 1     & 1     & 1     & 1     & 1 \\
          & 0.2   & 0,5344 & 0,5237 & 0,5320 & 0,9341 & 0,9629 & 1     & 1     & 1     & 1     & 0,9996 & 0,9996 \\
          & 0.3   & 0,1864 & 0,1637 & 0,6358 & 0,3756 & 0,4901 & 0,8602 & 0,8571 & 0,8575 & 0,9126 & 0,9063 & 0,9126 \\
    \midrule
    30    & 0.1   & 0,9923 & 0,9917 & 0,4932 & 1     & 1     & 1     & 1     & 1     & 1     & 1     & 1 \\
          & 0.2   & 0,1823 & 0,1581 & 0,2148 & 1     & 1     & 1     & 1     & 1     & 1     & 1     & 1 \\
          & 0.3   & 0,1658 & 0,1403 & 0,5935 & 0,4170 & 0,8556 & 0,6581 & 0,6485 & 0,6501 & 0,9670 & 0,9664 & 0,9689 \\
    \end{tabular}%
    }
  \label{tab:c_studentlambda5}%
\end{table}%

\begin{table}[htbp]
  \centering
  \caption{Correct detection rates with Student-t with 3 d.f. and  $\delta=10$.}
   \resizebox{15cm}{!} {
    \begin{tabular}{c|cccccccccccc}
           &     &   &   &     &     &   &   &   &   &   &   \\
          & $\lambda=0.1$ &       &       &       &       &       &       &       &       &       &       &  \\
    \midrule
    $p $    & $\alpha$ & MCD   & Adj MCD & Kurtosis & OGK   & COM   & RMDv1 & RMDv2 & RMDv3 & RMDv4 & RMDv5 & RMDv6 \\
    \midrule
    5     & 0.1   & 0,9900 & 0,9900 & 0,6600 & 1     & 1     & 1     & 1     & 1     & 1     & 1     & 1 \\
          & 0.2   & 0,5400 & 0,5400 & 0,4700 & 0,9600 & 1     & 1     & 1     & 1     & 1     & 1     & 1 \\
          & 0.3   & 0,0400 & 0,0400 & 0,8985 & 0,3803 & 0,7211 & 0,9900 & 0,9900 & 0,9900 & 1     & 1     & 0,9900 \\
    \midrule
    10    & 0.1   & 0,9300 & 0,9300 & 0,4900 & 1     & 1     & 1     & 1     & 1     & 1     & 1     & 1 \\
          & 0.2   & 0,1000 & 0,1000 & 0,2200 & 0,9300 & 1     & 1     & 1     & 1     & 1     & 1     & 1 \\
          & 0.3   & 0     & 0     & 0,7804 & 0,2200 & 0,7709 & 1     & 1     & 1     & 1     & 1     & 1 \\
    \midrule
    30    & 0.1   & 0,0200 & 0,0200 & 0,2594 & 1     & 1     & 1     & 1     & 1     & 1     & 1     & 1 \\
          & 0.2   & 0     & 0     & 0,3098 & 1     & 1     & 1     & 1     & 1     & 1     & 1     & 1 \\
          & 0.3   & 0,0001 & 0,0001 & 1     & 0,0003 & 0,7862 & 1     & 1     & 1     & 1     & 1     & 1 \\
    \midrule
    $p$     & $\alpha$ & $\lambda=1$ &       &       &       &       &       &       &       &       &       &  \\
    \midrule
    5     & 0.1   & 1     & 1     & 1     & 1     & 1     & 1     & 1     & 1     & 1     & 1     & 1 \\
          & 0.2   & 0,8890 & 0,8859 & 0,9580 & 1     & 1     & 1     & 1     & 1     & 1     & 1     & 1 \\
          & 0.3   & 0,2324 & 0,2127 & 0,9470 & 0,7598 & 0,9791 & 0,9998 & 0,9998 & 0,9998 & 1     & 1     & 1 \\
    \midrule
    10    & 0.1   & 1     & 1     & 0,9830 & 1     & 1     & 1     & 1     & 1     & 1     & 1     & 1 \\
          & 0.2   & 0,7050 & 0,6964 & 0,5019 & 1     & 1     & 1     & 1     & 1     & 1     & 1     & 1 \\
          & 0.3   & 0,2563 & 0,2314 & 0,6768 & 0,8453 & 0,9914 & 1     & 1     & 1     & 1     & 1     & 1 \\
    \midrule
    30    & 0.1   & 1     & 1     & 0,5614 & 1     & 1     & 1     & 1     & 1     & 1     & 1     & 1 \\
          & 0.2   & 0,2407 & 0,2166 & 0,2203 & 1     & 1     & 1     & 1     & 1     & 1     & 1     & 1 \\
          & 0.3   & 0,1640 & 0,1394 & 0,7130 & 0,9624 & 1     & 1     & 1     & 1     & 1     & 1     & 1 \\
    \end{tabular}%
    }
  \label{tab:c_studentlambda10}%
\end{table}%

\begin{table}[htbp]
  \centering
  \caption{False detection rates with Student-t with 3 d.f. and $\delta=5$.}
   \resizebox{15cm}{!} {
    \begin{tabular}{c|cccccccccccc}
           &     &   &   &     &     &   &   &   &   &   &   \\
     &       & $\lambda=0.1$ &       &       &       &       &       &       &       &       &       &  \\
    \midrule
    $p$     & $\alpha$ & MCD   & Adj MCD & Kurtosis & OGK   & COM   & RMDv1 & RMDv2 & RMDv3 & RMDv4 & RMDv5 & RMDv6 \\
    \midrule
    5     & 0.1   & 0,0901 & 0,0741 & 0,2020 & 0,1840 & 0,0641 & 0,1305 & 0,1293 & 0,1289 & 0,1155 & 0,1153 & 0,1208 \\
          & 0.2   & 0,1591 & 0,1381 & 0,3055 & 0,2014 & 0,0494 & 0,0741 & 0,0743 & 0,0748 & 0,0668 & 0,0656 & 0,0694 \\
          & 0.3   & 0,2237 & 0,1972 & 0,2046 & 0,4273 & 0,0654 & 0,0299 & 0,0300 & 0,0300 & 0,0298 & 0,0298 & 0,0310 \\
    \midrule
    10    & 0.1   & 0,1724 & 0,1528 & 0,3689 & 0,2149 & 0,0703 & 0,1876 & 0,1851 & 0,1874 & 0,1289 & 0,1259 & 0,1377 \\
          & 0.2   & 0,2460 & 0,2224 & 0,4709 & 0,2743 & 0,0534 & 0,0896 & 0,0882 & 0,0890 & 0,0643 & 0,0636 & 0,0687 \\
          & 0.3   & 0,2978 & 0,2723 & 0,3106 & 0,5624 & 0,0833 & 0,0411 & 0,0411 & 0,0410 & 0,0375 & 0,0364 & 0,0387 \\
    \midrule
    30    & 0.1   & 0,1775 & 0,1523 & 0,3154 & 0,2626 & 0,0717 & 0,3472 & 0,3450 & 0,3452 & 0,1563 & 0,1557 & 0,1581 \\
          & 0.2   & 0,2116 & 0,1831 & 0,4937 & 0,4232 & 0,0551 & 0,1459 & 0,1454 & 0,1452 & 0,0779 & 0,0777 & 0,0785 \\
          & 0.3   & 0,2562 & 0,2235 & 0,1669 & 0,8151 & 0,0991 & 0,0418 & 0,0418 & 0,0417 & 0,0401 & 0,0400 & 0,0446 \\
    \midrule
    & $\alpha$ & $\lambda=1$ &       &       &       &       &       &       &       &       &       &  \\
    \midrule
    5     & 0.1   & 0,0809 & 0,0662 & 0,2027 & 0,1903 & 0,0664 & 0,1361 & 0,1333 & 0,1346 & 0,1194 & 0,1183 & 0,1232 \\
          & 0.2   & 0,0619 & 0,0500 & 0,1846 & 0,1613 & 0,0513 & 0,0800 & 0,0799 & 0,0796 & 0,0711 & 0,0702 & 0,0727 \\
          & 0.3   & 0,1098 & 0,0951 & 0,1537 & 0,1340 & 0,0362 & 0,0375 & 0,0370 & 0,0372 & 0,0384 & 0,0380 & 0,0394 \\
    \midrule
    10    & 0.1   & 0,1209 & 0,1032 & 0,2985 & 0,2166 & 0,0769 & 0,1922 & 0,1910 & 0,1927 & 0,1414 & 0,1383 & 0,1497 \\
          & 0.2   & 0,1235 & 0,1073 & 0,2940 & 0,1805 & 0,0538 & 0,0986 & 0,0960 & 0,0962 & 0,0710 & 0,0701 & 0,0754 \\
          & 0.3   & 0,1705 & 0,1523 & 0,2096 & 0,1712 & 0,0360 & 0,0392 & 0,0394 & 0,0397 & 0,0361 & 0,0355 & 0,0375 \\
    \midrule
    30    & 0.1   & 0,1039 & 0,0819 & 0,1931 & 0,2587 & 0,0694 & 0,3466 & 0,3454 & 0,3460 & 0,1547 & 0,1540 & 0,1570 \\
          & 0.2   & 0,1514 & 0,1291 & 0,1888 & 0,2311 & 0,0557 & 0,1524 & 0,1519 & 0,1520 & 0,0781 & 0,0778 & 0,0788 \\
          & 0.3   & 0,1525 & 0,1304 & 0,1681 & 0,2248 & 0,0392 & 0,0445 & 0,0447 & 0,0444 & 0,0364 & 0,0363 & 0,0368 \\
    \end{tabular}%
    }
  \label{tab:f_studentlambda5}%
\end{table}%

\begin{table}[htbp]
  \centering
  \caption{False detection rates with Student-t with 3 d.f. and $\delta=10$.}
 \resizebox{15cm}{!} {
    \begin{tabular}{c|cccccccccccc}
           &     &   &   &     &     &   &   &   &   &   &   \\
     &       & $\lambda=0.1$ &       &       &       &       &       &       &       &       &       &  \\
    \midrule
   $ p $    & $\alpha$ & MCD   & Adj MCD & Kurtosis & OGK   & COM   & RMDv1 & RMDv2 & RMDv3 & RMDv4 & RMDv5 & RMDv6 \\
    \midrule
    5     & 0.1   & 0,0831 & 0,0669 & 0,2094 & 0,1955 & 0,0636 & 0,1378 & 0,1352 & 0,1368 & 0,1234 & 0,1222 & 0,1273 \\
          & 0.2   & 0,1090 & 0,0935 & 0,2544 & 0,1602 & 0,0442 & 0,0693 & 0,0690 & 0,0687 & 0,0629 & 0,0609 & 0,0647 \\
          & 0.3   & 0,2158 & 0,1904 & 0,1578 & 0,2988 & 0,0365 & 0,0330 & 0,0327 & 0,0331 & 0,0294 & 0,0295 & 0,0314 \\
    \midrule
    10    & 0.1   & 0,1272 & 0,1091 & 0,3392 & 0,2118 & 0,0737 & 0,1874 & 0,1851 & 0,1839 & 0,1316 & 0,1285 & 0,1387 \\
          & 0.2   & 0,2362 & 0,2148 & 0,4324 & 0,2050 & 0,0603 & 0,1033 & 0,1023 & 0,1013 & 0,0776 & 0,0761 & 0,0818 \\
          & 0.3   & 0,3045 & 0,2798 & 0,2388 & 0,4548 & 0,0346 & 0,0321 & 0,0320 & 0,0330 & 0,0253 & 0,0250 & 0,0277 \\
    \midrule
    30    & 0.1   & 0,1785 & 0,1533 & 0,3069 & 0,2618 & 0,0702 & 0,3474 & 0,3449 & 0,3457 & 0,1545 & 0,1530 & 0,1572 \\
          & 0.2   & 0,2129 & 0,1841 & 0,4603 & 0,2327 & 0,0529 & 0,1384 & 0,1382 & 0,1379 & 0,0647 & 0,0641 & 0,0659 \\
          & 0.3   & 0,2611 & 0,2282 & 0,1588 & 0,7518 & 0,0391 & 0,0387 & 0,0386 & 0,0387 & 0,0258 & 0,0258 & 0,0264 \\
    \midrule       
   & $\alpha$ & $\lambda=1$ &       &       &       &       &       &       &       &       &       &  \\
    \midrule
    5     & 0.1   & 0,0843 & 0,0660 & 0,1778 & 0,1854 & 0,0636 & 0,1368 & 0,1353 & 0,1361 & 0,1221 & 0,1213 & 0,1264 \\
          & 0.2   & 0,0454 & 0,0343 & 0,1855 & 0,1658 & 0,0506 & 0,0774 & 0,0779 & 0,0788 & 0,0670 & 0,0666 & 0,0684 \\
          & 0.3   & 0,0964 & 0,0818 & 0,1508 & 0,1287 & 0,0264 & 0,0287 & 0,0276 & 0,0274 & 0,0269 & 0,0269 & 0,0272 \\
    \midrule
    10    & 0.1   & 0,1151 & 0,0973 & 0,2771 & 0,2101 & 0,0716 & 0,1870 & 0,1859 & 0,1872 & 0,1325 & 0,1295 & 0,1403 \\
          & 0.2   & 0,0892 & 0,0768 & 0,2694 & 0,1759 & 0,0531 & 0,0896 & 0,0872 & 0,0887 & 0,0641 & 0,0635 & 0,0682 \\
          & 0.3   & 0,1494 & 0,1337 & 0,2122 & 0,1437 & 0,0328 & 0,0371 & 0,0360 & 0,0360 & 0,0305 & 0,0300 & 0,0318 \\
    \midrule
    30    & 0.1   & 0,1027 & 0,0803 & 0,1994 & 0,2635 & 0,0710 & 0,3423 & 0,3417 & 0,3420 & 0,1523 & 0,1509 & 0,1552 \\
          & 0.2   & 0,1451 & 0,1233 & 0,1937 & 0,2297 & 0,0551 & 0,1453 & 0,1453 & 0,1455 & 0,0685 & 0,0679 & 0,0699 \\
          & 0.3   & 0,1501 & 0,1279 & 0,1633 & 0,1997 & 0,0409 & 0,0390 & 0,0388 & 0,0389 & 0,0244 & 0,0243 & 0,0250 \\
    \end{tabular}%
    }
  \label{tab:f_studentlambda10}%
\end{table}%

\newpage

\subsection{Multivariate Exponential distribution}

Table \ref{tab:0contame} shows the false detection rates when there is no contamination.
Table \ref{tab:c_exponential} shows the correct detection rates (c) and Table \ref{tab:f_exponential} the false detection rates (f), for each method, corresponding to the simulations with multivariate Exponential distribution, for contamination levels $\alpha=0.1,0.2,0.3$, dimension $p=5,10,30$ and distance of the outliers $\delta=5,10$.

\begin{table}[htbp]
  \centering
  \caption{False detection rates with Exponential distribution $\alpha=0$.}
  \resizebox{15cm}{!} {
    \begin{tabular}{c|ccccccccccc}
           &     &   &   &     &     &   &   &   &   &   &   \\
   $ p$     & MCD   & Adj MCD & Kurtosis & OGK   & COM   & RMDv1 & RMDv2 & RMDv3 & RMDv4 & RMDv5 & RMDv6 \\
    \midrule
    5     & 0,15070 & 0,13310 & 0,29900 & 0,26540 & 0,21340 & 0,16060 & 0,15850 & 0,16090 & 0,11820 & 0,11580 & 0,11280 \\
    10    & 0,18710 & 0,16890 & 0,36000 & 0,26580 & 0,38930 & 0,13638 & 0,15970 & 0,15910 & 0,13454 & 0,13920 & 0,13609 \\
    30    & 0,14292 & 0,11952 & 0,13558 & 0,27938 & 0,44708 & 0,13324 & 0,13266 & 0,13246 & 0,18204 & 0,18092 & 0,18464 \\
    \end{tabular}%
    }
  \label{tab:0contame}%
\end{table}%

\begin{table}[htbp]
\centering
  \caption{Correct detection rates with Exponential distribution.}
   \resizebox{15cm}{!} {
    \begin{tabular}{c|cccccccccccc}
           &     &   &   &     &     &   &   &   &   &   &   \\
          &       & $\delta=5$ &       &       &       &       &       &       &       &       &       &  \\
    \midrule
    $p$     & $\alpha$ & MCD   & Adj MCD & Kurtosis & OGK   & COM   & RMDv1 & RMDv2 & RMDv3 & RMDv4 & RMDv5 & RMDv6 \\
    \midrule
    5     & 0.1   & 1     & 1     & 1     & 1     & 1     & 1     & 1     & 1     & 1     & 1     & 1 \\
          & 0.2   & 0,8578 & 0,8486 & 0,9602 & 0,9654 & 0,9975 & 1     & 1     & 1     & 0,9993 & 0,9993 & 0,9993 \\
          & 0.3   & 0,1955 & 0,1664 & 0,8336 & 0,5792 & 0,8735 & 0,8935 & 0,8947 & 0,8938 & 0,8740 & 0,8698 & 0,8755 \\
    \midrule
    10    & 0.1   & 0,9983 & 0,9976 & 0,9974 & 0,9983 & 0,9975 & 0,9993 & 0,9993 & 0,9993 & 0,9995 & 0,9995 & 0,9999 \\
          & 0.2   & 0,9671 & 0,9649 & 0,9893 & 0,9984 & 0,9923 & 0,9992 & 0,9992 & 0,9992 & 0,9984 & 0,9984 & 0,9986 \\
          & 0.3   & 0,7821 & 0,7792 & 0,9847 & 0,9887 & 0,9832 & 0,9957 & 0,9958 & 0,9957 & 0,9951 & 0,9951 & 0,9958 \\
    \midrule
    30    & 0.1   & 1     & 1     & 1     & 1     & 1     & 1     & 1     & 1     & 1     & 1     & 1 \\
          & 0.2   & 0,9985 & 0,9985 & 0,9987 & 1     & 1     & 1     & 1     & 1     & 1     & 1     & 1 \\
          & 0.3   & 0,7829 & 0,7550 & 1     & 1     & 1     & 1     & 1     & 1     & 1     & 1     & 1 \\
    \midrule
    $p$     & $\alpha$ & $\delta=10$ &       &       &       &       &       &       &       &       &       &  \\
    \midrule
    5     & 0.1   & 0,9991 & 0,9970 & 1     & 1     & 0,9991 & 0,9991 & 0,9991 & 0,9991 & 1     & 1     & 1 \\
          & 0.2   & 0,9770 & 0,9733 & 0,9991 & 0,9991 & 0,9927 & 0,9961 & 0,9961 & 0,9961 & 0,9952 & 0,9952 & 0,9997 \\
          & 0.3   & 0,8185 & 0,7952 & 0,9900 & 0,9966 & 0,9853 & 0,9916 & 0,9919 & 0,9919 & 0,9906 & 0,9903 & 0,9911 \\
    \midrule
    10    & 0.1   & 1     & 1     & 1     & 1     & 1     & 1     & 1     & 1     & 1     & 1     & 1 \\
          & 0.2   & 0,9756 & 0,9756 & 1     & 1     & 1     & 1     & 1     & 1     & 1     & 1     & 1 \\
          & 0.3   & 0,7905 & 0,7902 & 1     & 1     & 1     & 1     & 1     & 1     & 1     & 1     & 1 \\
    \midrule
    30    & 0.1   & 1     & 1     & 1     & 1     & 1     & 1     & 1     & 1     & 1     & 1     & 1 \\
          & 0.2   & 0,9987 & 0,9987 & 1     & 1     & 1     & 1     & 1     & 1     & 1     & 1     & 1 \\
          & 0.3   & 0,7719 & 0,7601 & 1     & 1     & 1     & 1     & 1     & 1     & 1     & 1     & 1 \\
    \end{tabular}%
    }
  \label{tab:c_exponential}%
\end{table}%
%
%
\begin{table}[htbp]
  \centering
  \caption{False detection rates with Exponential distribution.}
  \resizebox{15cm}{!} {
    \begin{tabular}{c|cccccccccccc}
           &     &   &   &     &     &   &   &   &   &   &   \\
          &       & $\delta=5$ &       &       &       &       &       &       &       &       &       &  \\
    \midrule
    $p$     & $\alpha$ & MCD   & Adj MCD & Kurtosis & OGK   & COM   & RMDv1 & RMDv2 & RMDv3 & RMDv4 & RMDv5 & RMDv6 \\
    \midrule
    5     & 0.1   & 0,0339 & 0,0198 & 0,0341 & 0,0636 & 0,0033 & 0,0084 & 0,0073 & 0,0069 & 0,0075 & 0,0071 & 0,0077 \\
          & 0.2   & 0,0110 & 0,0055 & 0,0347 & 0,0507 & 0,0017 & 0,0014 & 0,0012 & 0,0012 & 0,0010 & 0,0009 & 0,0012 \\
          & 0.3   & 0,0374 & 0,0260 & 0,0379 & 0,0383 & 0,0001 & 0,0000 & 0,0000 & 0,0000 & 0,0000 & 0,0000 & 0,0000 \\
    \midrule
    10    & 0.1   & 0,1194 & 0,1018 & 0,3285 & 0,2299 & 0,0683 & 0,2380 & 0,2362 & 0,2372 & 0,2302 & 0,2258 & 0,2442 \\
          & 0.2   & 0,0315 & 0,0235 & 0,2634 & 0,1838 & 0,0466 & 0,1176 & 0,1183 & 0,1182 & 0,1338 & 0,1302 & 0,1479 \\
          & 0.3   & 0,0021 & 0,0016 & 0,1890 & 0,1445 & 0,0258 & 0,0575 & 0,0576 & 0,0576 & 0,0722 & 0,0681 & 0,0803 \\
    \midrule
    30    & 0.1   & 0,0887 & 0,0639 & 0,1276 & 0,2371 & 0,0335 & 0,4021 & 0,4020 & 0,4015 & 0,2726 & 0,2697 & 0,2781 \\
          & 0.2   & 0,0239 & 0,0099 & 0,1390 & 0,2022 & 0,0202 & 0,1751 & 0,1750 & 0,1754 & 0,1531 & 0,1504 & 0,1583 \\
          & 0.3   & 0,0001 & 0,0000 & 0,1196 & 0,1734 & 0,0111 & 0,0525 & 0,0528 & 0,0524 & 0,0551 & 0,0541 & 0,0580 \\
    \midrule
   $ p $    & $\alpha$ & $\delta=10$ &       &       &       &       &       &       &       &       &       &  \\
    \midrule
    5     & 0.1   & 0,0842 & 0,0679 & 0,2846 & 0,2198 & 0,0799 & 0,1567 & 0,1572 & 0,1557 & 0,1659 & 0,1631 & 0,1708 \\
          & 0.2   & 0,0271 & 0,0176 & 0,2408 & 0,1824 & 0,0551 & 0,0743 & 0,0752 & 0,0754 & 0,1015 & 0,0994 & 0,1039 \\
          & 0.3   & 0,0011 & 0,0006 & 0,1969 & 0,1451 & 0,0288 & 0,0302 & 0,0299 & 0,0289 & 0,0460 & 0,0443 & 0,0470 \\
    \midrule
    10    & 0.1   & 0,1148 & 0,0961 & 0,3201 & 0,2284 & 0,0620 & 0,2103 & 0,2121 & 0,2131 & 0,2015 & 0,1963 & 0,2161 \\
          & 0.2   & 0,0357 & 0,0261 & 0,2647 & 0,1841 & 0,0404 & 0,0895 & 0,0906 & 0,0890 & 0,0942 & 0,0912 & 0,1032 \\
          & 0.3   & 0,0028 & 0,0022 & 0,1873 & 0,1474 & 0,0240 & 0,0298 & 0,0293 & 0,0309 & 0,0411 & 0,0388 & 0,0463 \\
    \midrule
    30    & 0.1   & 0,0863 & 0,0621 & 0,1343 & 0,2416 & 0,0334 & 0,3536 & 0,3530 & 0,3536 & 0,2394 & 0,2371 & 0,2450 \\
          & 0.2   & 0,0253 & 0,0116 & 0,1179 & 0,2043 & 0,0213 & 0,1053 & 0,1056 & 0,1058 & 0,0913 & 0,0899 & 0,0944 \\
          & 0.3   & 0,0000 & 0,0000 & 0,1240 & 0,1662 & 0,0108 & 0,0127 & 0,0128 & 0,0130 & 0,0146 & 0,0144 & 0,0156 \\
    \end{tabular}%
    }
  \label{tab:f_exponential}%
\end{table}%

\newpage


\subsection{Figures}\label{appfigures}

The following are figures corresponding to the real dataset example.\\

\begin{figure}[H]
     \centering
     \includegraphics[width=1.1\textwidth]{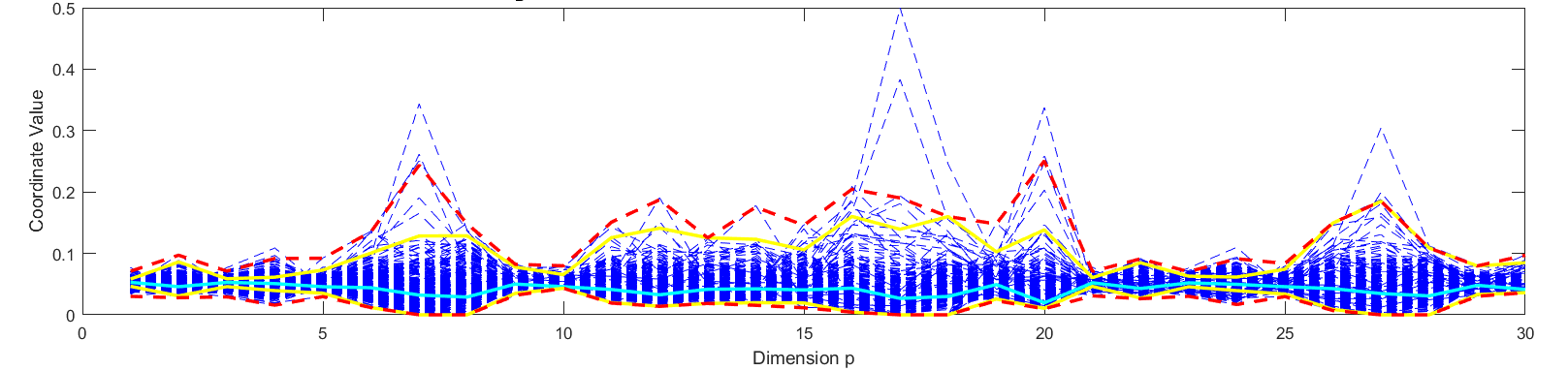}
     \caption{{\footnotesize Standardized data with the ``multivariate boxplot''.}}            
     \label{databoxplot}
\end{figure}

\begin{figure}[H]
     \centering
     \includegraphics[scale=0.3]{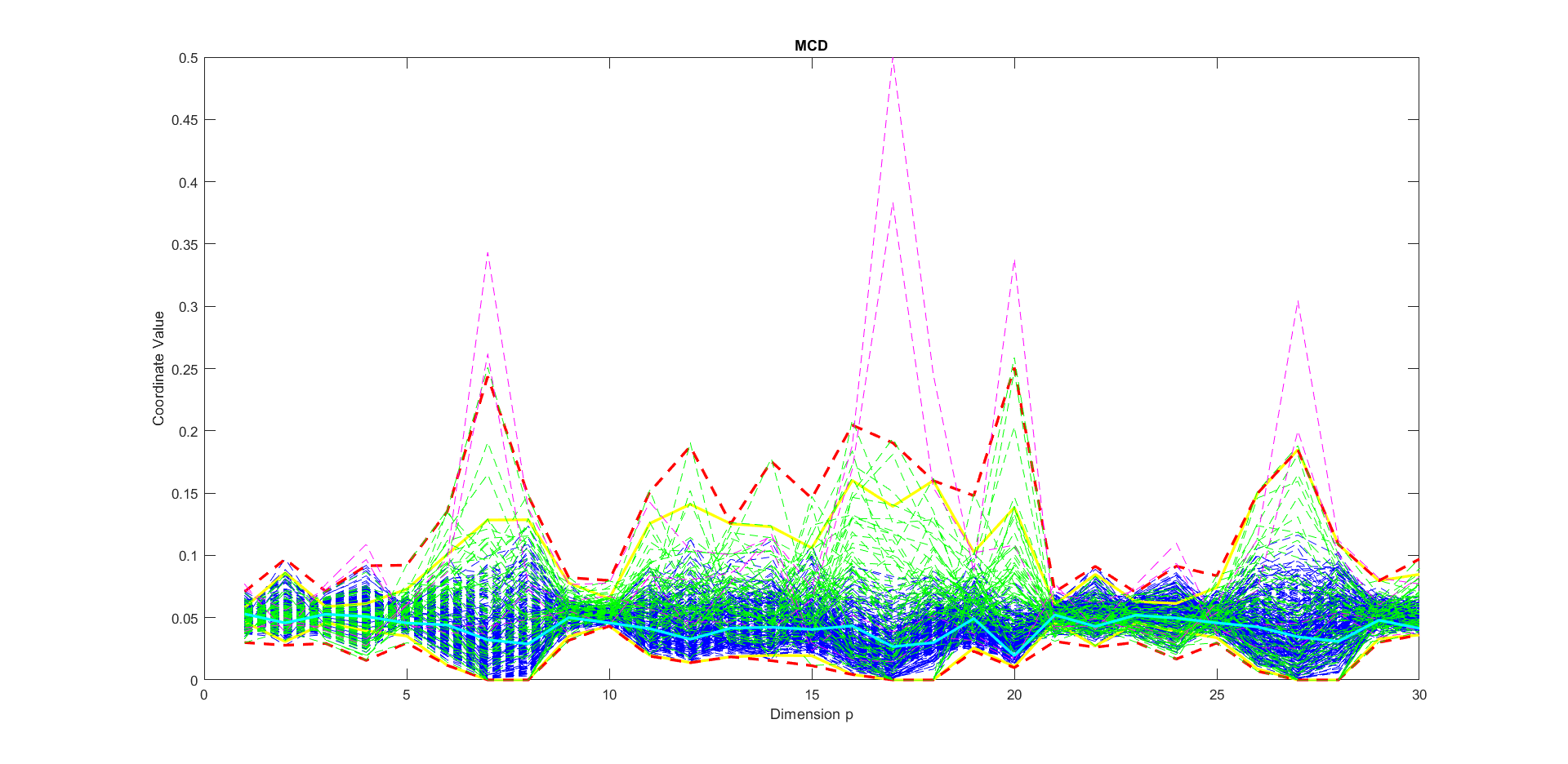}
     \caption{{\footnotesize Detected outliers by MCD.}}            
     \label{MCD}
\end{figure}

\begin{figure}[H]
     \centering
     \includegraphics[scale=0.3]{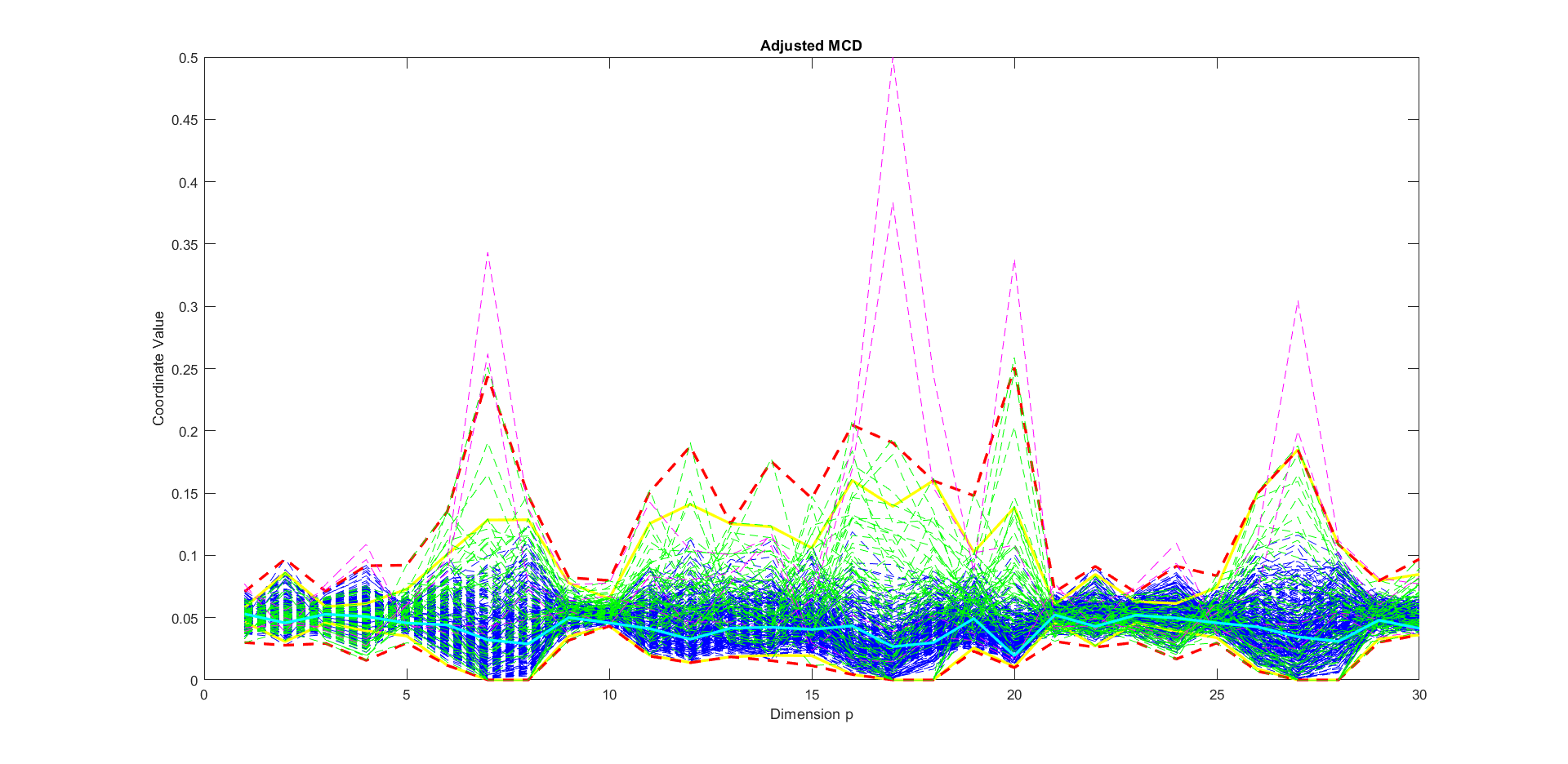}
     \caption{{\footnotesize Detected outliers by Adjusted MCD.}}            
     \label{AdjMCD}
\end{figure}

\begin{figure}[H]
     \centering
     \includegraphics[scale=0.3]{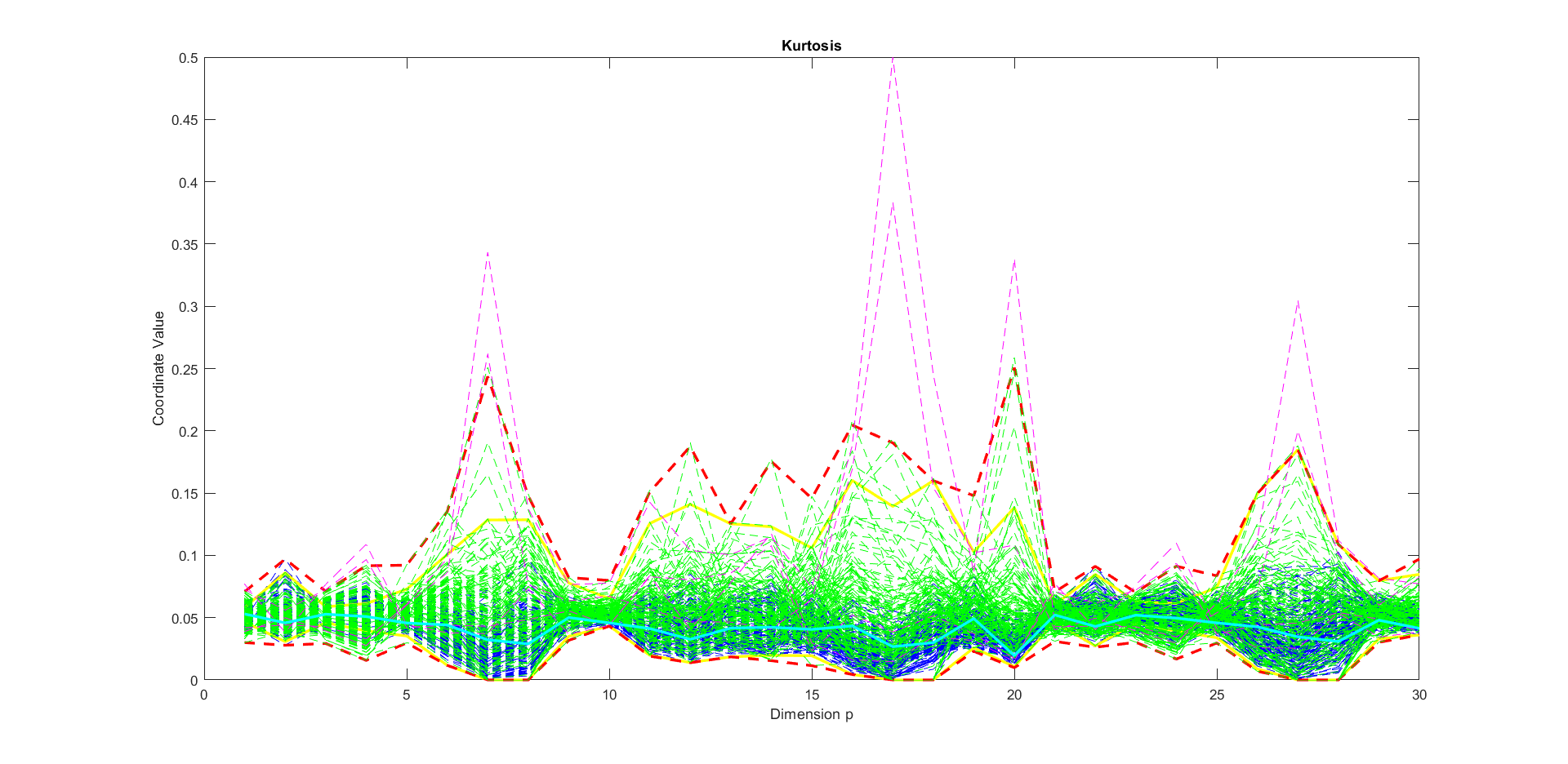}
     \caption{{\footnotesize Detected outliers by Kurtosis.}}            
     \label{Kurtosis}
\end{figure}

\begin{figure}[H]
     \centering
     \includegraphics[scale=0.3]{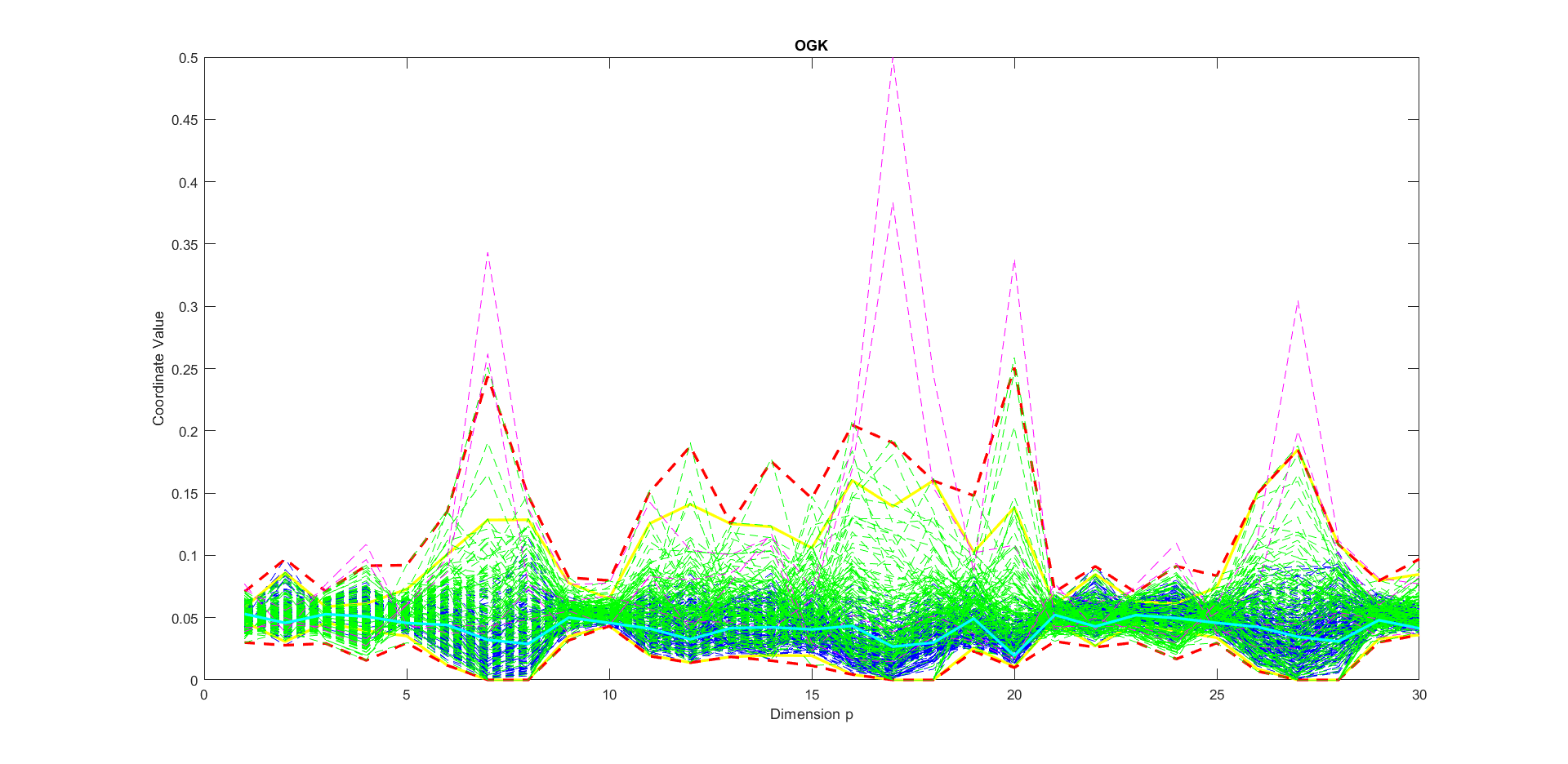}
     \caption{{\footnotesize Detected outliers by OGK.}}            
     \label{OGK}
\end{figure}

\begin{figure}[H]
     \centering
     \includegraphics[scale=0.3]{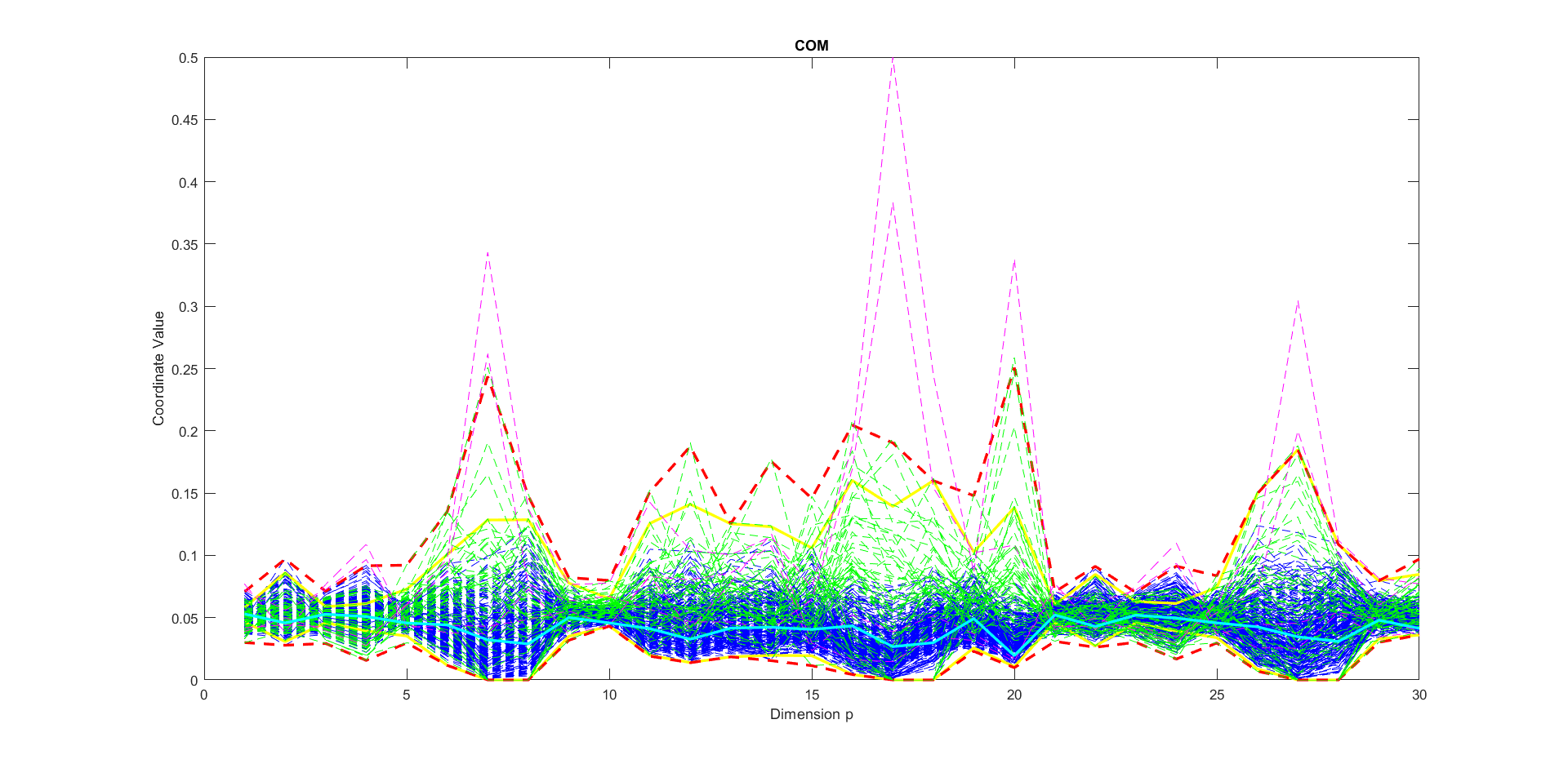}
     \caption{{\footnotesize Detected outliers by COM.}}            
     \label{COM}
\end{figure}

\begin{figure}[H]
     \centering
     \includegraphics[scale=0.3]{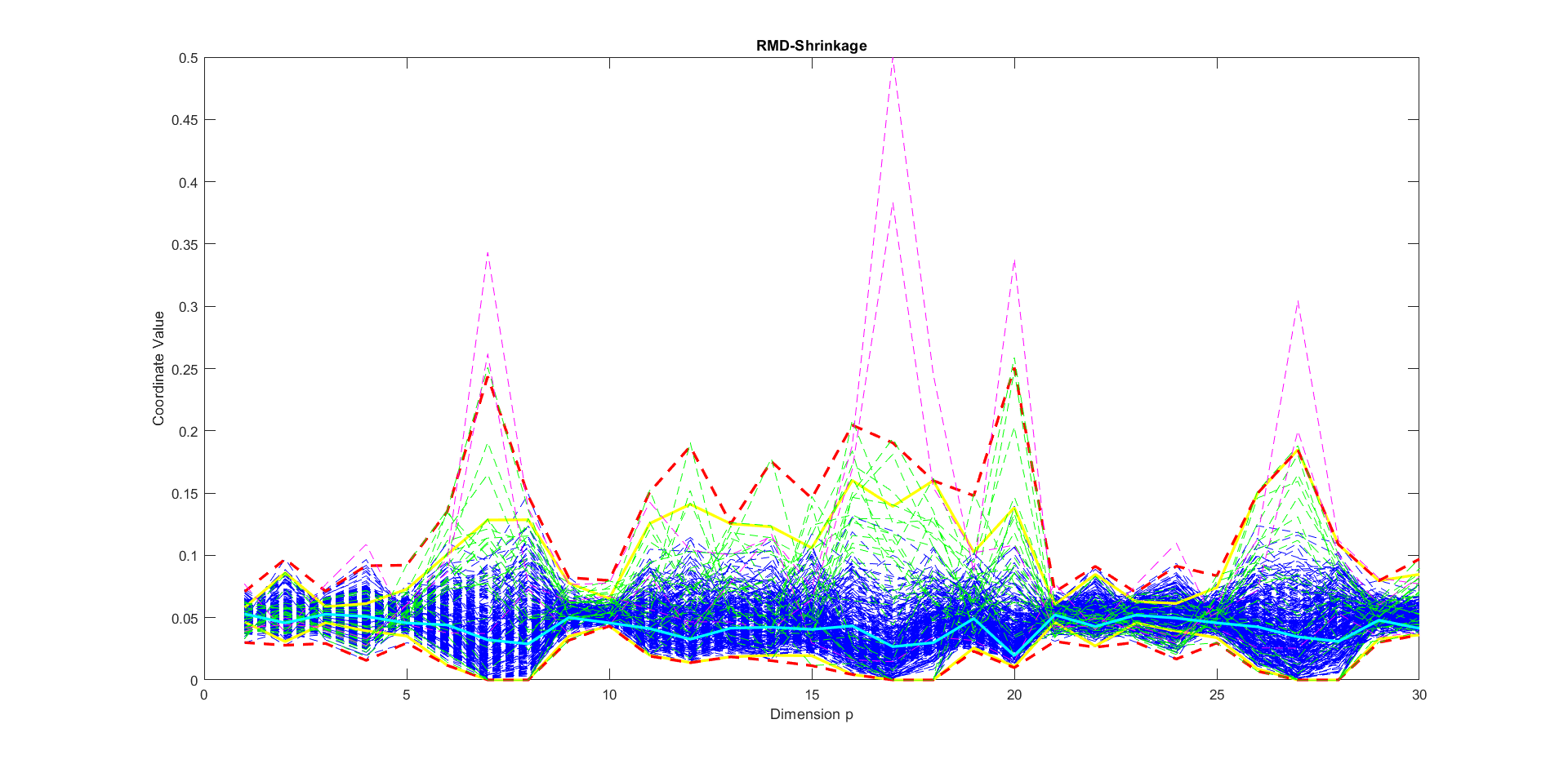}
     \caption{{\footnotesize Detected outliers by RMD-Shrinkage.}}            
     \label{RMD-Shrinkage}
\end{figure}

\begin{figure}[H]
     \centering
     \includegraphics[width=1.1\textwidth]{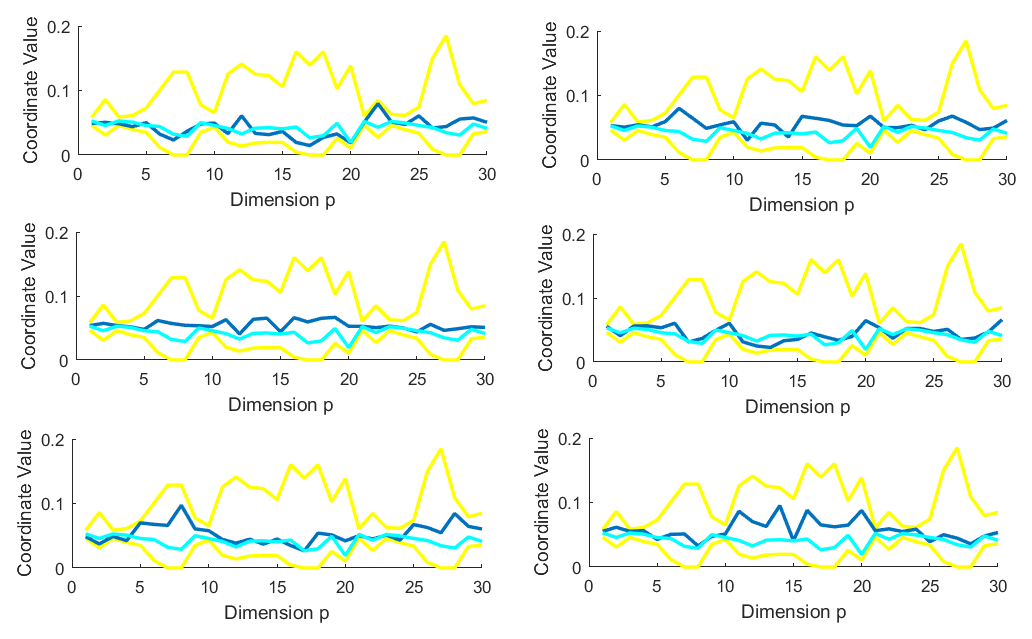}
     \caption{{\footnotesize Some of our competitors detected outliers belonging to the $50\%$ of the most central
data.}}            
     \label{outliersrealdata}
\end{figure}

\begin{figure}[H]
     \centering
     \includegraphics[width=1.3\textwidth]{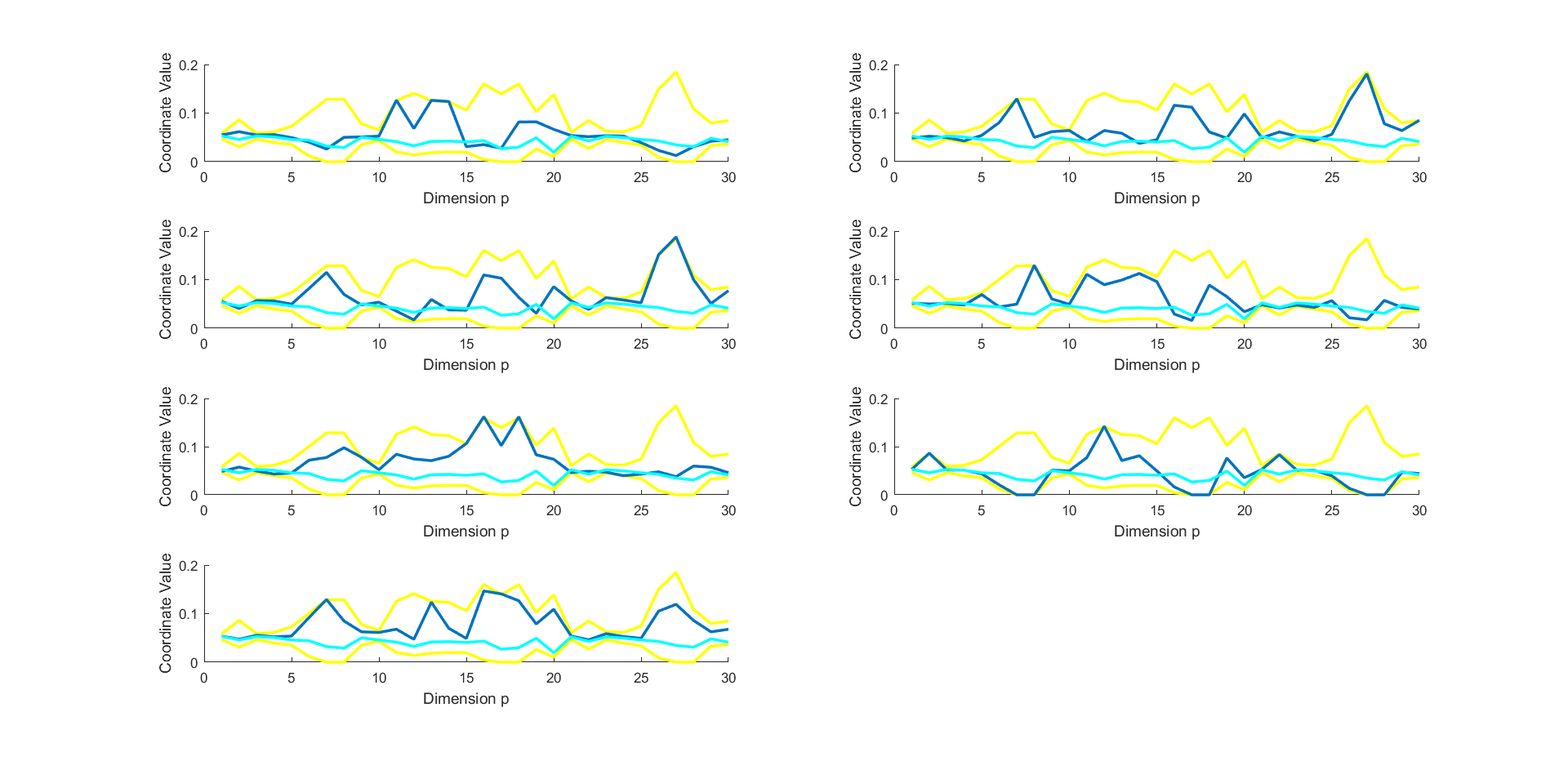}
     \caption{{\footnotesize RMD-Shrinkage detected outliers that belong to the $50\%$ of the most central data.}}            
     \label{outliersRMDv62}
\end{figure}

\bibliography{bbib}

\end{document}